\documentclass[journal]{IEEEtran}
\IEEEoverridecommandlockouts
\usepackage{url}
\usepackage[dvipdfmx]{graphicx} 
\usepackage{bmpsize}
\usepackage{mathtools}
\usepackage{amsmath}
\usepackage{textcomp}
\usepackage[caption=false,font=footnotesize,labelfont=sf,textfont=sf]{subfig}
\usepackage[belowskip=-7pt,aboveskip=3pt]{caption}
\usepackage{listings}
\usepackage{color}
\usepackage{hyperref}
\usepackage{lipsum,multicol}
\usepackage[linesnumbered,algoruled]{algorithm2e}
\usepackage{float}
\usepackage{tikz}
\definecolor{lightgray}{rgb}{.9,.9,.9}
\definecolor{darkgray}{rgb}{.4,.4,.4}
\definecolor{purple}{rgb}{0.65, 0.12, 0.82}
\SetAlFnt{\small}
\SetKwComment{tcp}{\small // }{}%
\SetCommentSty{small}
\newcommand\tab[1][0.3cm]{\hspace*{#1}}
\newcommand\stab[1][0.7cm]{\hspace*{#1}}
\newcommand\mtab[1][1.1cm]{\hspace*{#1}}
\newcommand\ltab[1][1.5cm]{\hspace*{#1}}

\lstdefinelanguage{JavaScript}{
  keywords={jcond, jdata, jsync, jasync, var, typeof, new, true, false, catch, logger, char, struct, broadcaster, inflow, outflow, return, null, catch, switch, if, in, while, do, else, case, break},
  keywordstyle=\color{blue}\bfseries,
  ndkeywords={class, var, export, boolean, throw, implements, import, this},
  ndkeywordstyle=\color{green}\bfseries,
  identifierstyle=\color{black},
  sensitive=false,
  comment=[l]{//},
  morecomment=[s]{/*}{*/},
  commentstyle=\color{purple}\ttfamily,
  stringstyle=\color{red}\ttfamily,
  morestring=[b]',
  morestring=[b]"
}

\begin{document}

\title{Multi-Point Synchronization for Fog-Controlled Internet of Things
}
\author{Richard~Olaniyan and~Muthucumaru~Maheswaran
\thanks{Research supported by Petroleum Technology Development Fund (PTDF), Nigeria and a Natural Sciences and Engineering Research Council of Canada (NSERC) Discovery Grant.}
\thanks{\textcopyright 2019 IEEE.  Personal use of this material is permitted.  Permission from IEEE must be obtained for all other uses, in any current or future media, including reprinting/republishing this material for advertising or promotional purposes, creating new collective works, for resale or redistribution to servers or lists, or reuse of any copyrighted component of this work in other works.}

}

\maketitle

\begin{abstract}
This paper presents a fog-resident controller architecture for synchronizing the operations of large collections of Internet of Things (IoT) such as drones, Internet of Vehicles, etc.
Synchronization in IoT is grouped into different classes, use cases identified and  multi-point synchronous scheduling algorithms are developed to schedule tasks with varying timing requirements; strict (synchronous) and relaxed (asynchronous and local) onto a bunch of worker nodes that are coordinated by a fog resident controller in the presence of disconnections and worker failures.  The algorithms use time-based or component-based redundancy to cope with failures and embed a publish-subscribe message update scheme to reduce the message overhead at the controller as the number of workers increase. The performance of the algorithms are evaluated using trace-driven experiments and practicability is shown by implementing the time-based redundancy synchronous scheduling algorithm in JAMScript -- a polyglot programming platform for Cloud of Things and report initial findings.\\


\end{abstract}

\section{Introduction}\label{sec1}

\IEEEPARstart{I}{oT} come in many shapes and sizes. Because each IoT component will be handling a part of a problem domain, cooperative computing among the IoT components is essential for solving many interesting problems~\cite{abedin2015fog}. For example, consider a situation where a swarm of drones is required to take a snapshot of a search area. A synchronization mechanism is required to ensure that the autonomous drones capture a specified part of the search area at the same time. If the drones do not follow strict timing alignments, it will lead to a poor reconstruction of the search area. As another example take an autonomous car, the car could form a swarm with other cars and road side units. To drive successfully, the car needs assurance that data is collected within a short time interval by the sensing and processing components of the swarm. The car thus needs support from the swarm to ensure that its tasks are run simultaneously across the different constituents. 

Cooperating IoT need to run tasks of the application in a coordinated manner. One such coordination is running the tasks at the same time. For instance drones lifting a workload need to exert the force at the same time so the effort adds up. To run the tasks at the same time, there is a  need for synchronization primitives that would launch the tasks with the start times lined up at all IoT devices. Just having synchronized clocks~\cite{wu2011clock,guo2016psync} is not sufficient because even with an earlier agreed start time, the IoT devices could fail, disconnect from the swarm~\cite{lee2014swarm}, or be busy with a prior task.  With tighter synchronization schemes, fine-grained tasks can be run across the cooperating IoT. 

In this paper, a controller-based scheme for synchronizing the execution of the tasks across a cooperating set of IoT devices is presented. With a controller-based scheme, one of the issues is the location of the controller itself. The recent emergence of fog computing as a major complementary technology to IoT is making it an ideal candidate to host our controller. Fogs are placed closer to the IoT devices they serve, placing the controller on a fog makes it accessible with the lowest possible latency from all IoT devices in the cooperating set~\cite{bonomi2012fog}.

The primary focus of this paper is on developing such synchronization schemes that guarantees the desired \textit{quality of synchronization}, $QoS_{ync}$ subject to the conditions under which synchronization is deemed successful. The synchronization scheme has to ensure that the constituent tasks of an application (asynchronous, synchronous and local tasks) are scheduled onto a bunch of cooperating devices while ensuring that the $QoS_{ync}$ is met.

The major contributions of this work are as follows:

\begin{enumerate}
\item A hierarchical system with a controller-worker model for running tasks with strict coordination among cooperating IoT components is presented.

\item Two redundancy-based dynamic synchronous scheduling algorithms with different synchrony requirements to handle synchronous, asynchronous and local tasks are developed.

\item The practicability of the synchronous scheduling scheme is shown by implementing the time-based redundancy algorithm in JAMScript -- ``a programming language and framework developed for Cloud of Things~\cite{wenger2016programming}''.
\end{enumerate}

Hereafter, IoT devices are referred to as \textit{workers} and {\em sync} is used to refer to ``synchronization'' or ``synchronous''. The rest of this paper is structured as follows. A background and motivation for this paper is given in Section~\ref{sec2}. The system architecture comprising of the node, application and task models is presented in Section~\ref{sec3}. The algorithms and experimental results are presented in Sections~\ref{sec4} and ~\ref{sec5} respectively.  Section~\ref{sec6} provides details on implementation and the observations made. The related work is given in Section~\ref{sec7}.

\section{Background and Motivation}\label{sec2}

Synchronization is a well studied topic in computer science. In gang scheduling~\cite{karatza2006scheduling} and coscheduling ~\cite{sobalvarro1998dynamic} where tasks of a parallel job are scheduled and executed at the same dedicated node, synchronization is achieved by using busy waiting. In parallel computing models such as the bulk synchronous parallel (BSP) model~\cite{valiant1990bridging} where the execution model is broken into computation and communication super steps, synchronization is achieved by using barriers. Barrier synchronization involves processes stopping at the barrier until all other processes reach the barrier~\cite{khayyat2013mizan, jakovits2012stratus}. Thus, faster processes have to wait for slower processes, with the bottleneck being the slowest process. Time slotting has been adopted as a way of achieving synchronization in wireless sensor networks~\cite{vogli2015fast, ozil2007time}. Dedicated synchronization time slots are chosen and devices attempt to perform synchronization tasks only in the dedicated time slots. Time slotting suffers from stragglers as with barrier synchronization. Slow workers will miss the dedicated synchronization slot and will result in the synchronization failing.


Synchronization schemes in distributed systems cannot be directly applied to IoT due to the following factors. (i) Node connectivity could be highly unstable in IoT due to mobility and disconnections unlike distributed systems where stable connection exists among nodes. Synchronization schemes in distributed systems are developed based on this assumption and are thus not suitable for highly dynamic systems. (ii) The network topology rapidly changes due to nodes joining and leaving in IoT unlike distributed systems. (iii) There is an interaction with physical things that have real-time window constraints in IoT unlike traditional distributed systems which do not necessarily affect real world systems. (iv) Nodes in IoT could be highly heterogeneous e.g., sensors, mobile phones, cars etc., unlike distributed systems where nodes have similar characteristics.


\subsection{Definition and Taxonomy of Synchronization in IoT}\label{definition}

Synchronization in IoT in the scope of this work is defined as the coordination of a set of IoT nodes to harmonize on the execution of a task at the same point or at multiple points in time. 

\begin{figure*}[htbp]
\centering
\includegraphics[width=6.2in]{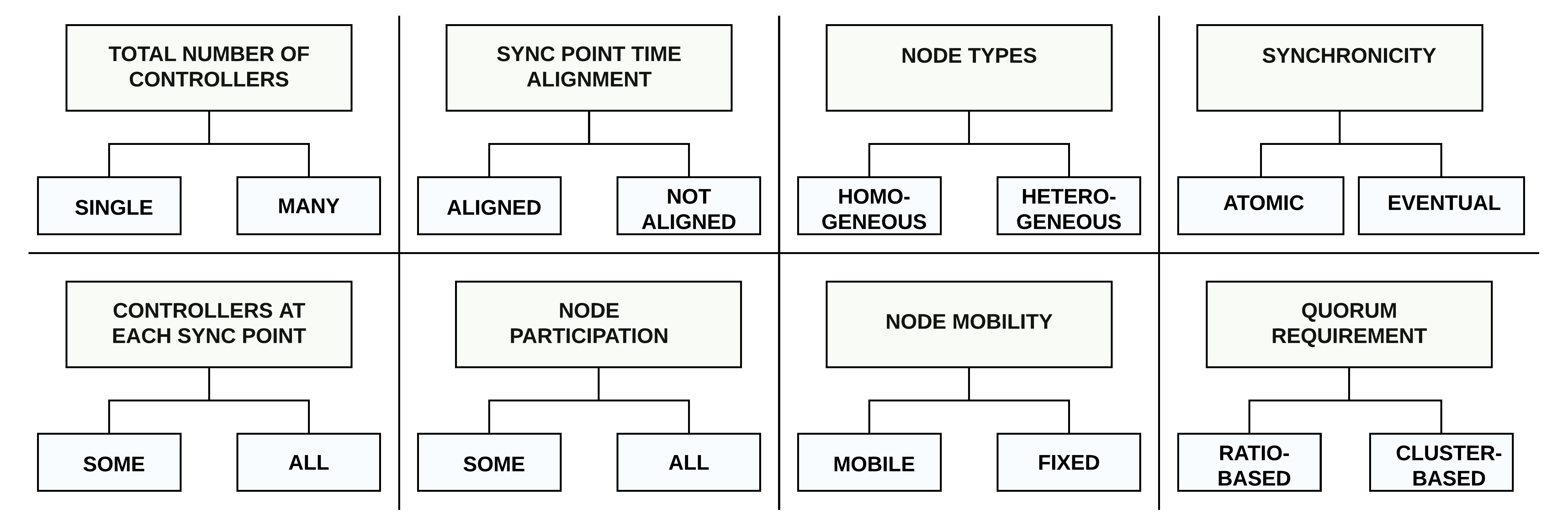}
\caption{Taxonomy of synchronization in IoT.}
\label{taxonomy}
\end{figure*}

Classification of synchronization in IoT is done based on unique attributes and requirements of synchronization in IoT. The taxonomy of synchronization in IoT is shown in Fig.~\ref{taxonomy} and the classification is as follows.

\subsubsection{Total number of controllers}

The number of controllers in an IoT system could be single or many depending on the scope of the system. In the case of a single controller, all workers must be connected to the controller to be part of the system. In a multi-controller system, workers connect to the controller that is closest to them or assigned to their physical vicinity. Workers can change controller membership as they move around.

\subsubsection{Number of controllers at synchronization point}

Synchronization could be at a global scale or local scale. In a global synchronization, all the controllers in the system must ensure that the devices under them synchronize their activities on the given sync task. In a local synchronization, only a subset of controllers that are affected by the synchronization are involved in the synchronization process. 

\subsubsection{Synchronization point time alignment}

All the controllers participating in the synchronization can either orchestrate the devices under them to start the execution of the sync task at the same time or at different times. Thus, a single controller system can only permit an aligned sync point time while a multi-controller system can permit both an aligned and a non-aligned sync point time.

\subsubsection{Node participation}

Some synchronization tasks require either all or some specified number of the devices connected to a controller to partake in the synchronization process. The participation of devices is dependent on their availability to run the sync task and the requirements of the sync task itself.

\subsubsection{Node types}

Synchronization in IoT is affected by node types. Nodes could be homogeneous, having similar attributes or heterogeneous, having different attributes.

\subsubsection{Node mobility}

Mobility is an important factor affecting synchronization in IoT. Nodes could be mobile, such as hand-held devices, cars etc, or fixed such as nodes placed on lamp posts.



\subsubsection{Quorum requirement}\label{qcheck}

The conditions at which synchronization can occur are either having at least the required fraction of nodes available to run the synchronous task (\textit{ratio-based quorum}) or having the desired representation per group of clustered nodes (\textit{cluster-based quorum}). The $QoS_{ync}$ is defined by the quorum requirement conditions. 

\subsubsection{Synchronicity}

In atomic synchronization, a sync task is executed or not depending on the quorum success or failure. In eventual synchronization, the sync task is run regardless of whether the nodes are synchronized, synchronicity is expected to increase with time.

\subsection{Synchronization Use Cases}

The use cases for synchronization in IoT are identifies as follows:

\subsubsection{Capacity pool} 

IoT and other smart devices are usually limited in their computing and sensing capabilities. Thus, there is a need for cooperation among several devices to solve a much bigger problem than can be solved by an individual device. The devices must pool their resources together in a coordinated manner in order to be successful. An example is autonomous drones pooling their lifting capacities to carry a large package from one point to another. 
    
\subsubsection{Data capture synchrony}

Data synchronization~\cite{liu2015} is an important problem in IoT  data acquisition. Using synchronous tasks for capturing data allows us to precisely control the relative timing relations among the data points. In data capture synchrony, different subset of nodes can be made to synchronize at different points in time thus achieving some level of data ordering.
    
\subsubsection{Resource usage synchrony} 

IoT devices consume resources (e.g. power) for their operation; therefore, there is a need for subset 
synchronization to sequence the operating order of the IoT devices to minimize the maximum resource usage profile. 
Take for example a smart lighting system consisting of a large number of bulbs, to light up (cover) a particular area, only a subset of the bulbs need to be turned on at the same time. Synchronization can be used to incrementally change the lighting intensity or maintain a constant lighting intensity.

\subsection{Example Deployment Scenarios}
Three application scenarios where synchronization among devices is of high importance are given.

\subsubsection{Drone Delivery System}

Drone transportation~\cite{chalupnivckova2014use} is becoming increasingly popular and lots of research and industrial effort is going into using drones as a means of transporting humans, goods (e.g. Amazon and UPS), vaccines and other products. In goods transportation, rather than use a large drone, smaller drones could be deployed and coordinated to carry heavy goods from one facility to another.

Drone transportation is an example of capacity pooling where the required number of drones is pooled from the available drones with component-based redundancy included to cater for failures.
Time, location and redundancy are important factors in drone transportation. A certain number of drones are needed at the start location before a lift could commence. Predictive models can be used to estimate the availabilities of the drones at different locations and times which can be used to improve the overall transportation schedule.  

\subsubsection{Smart Cars and Self Driving}

A smart (autonomous) car carries enormous amount of computing, storage and sensing capacities~\cite{glancy2015autonomous}. With fast wireless networks and fog computing, smart cars can share their capacities with other cars or with the road side infrastructure and vice-versa. Such a swarm of cars will have significantly augmented capabilities; that is, a smart car could have more capabilities (i.e., higher level autonomy~\cite{hasan2011intelligent}) on a smart highway (SH) than what is capable of on a normal highway. 

Smart cars need up-to-date information about the status of the road (presence of other cars and the road condition) and other driving conditions (such as weather) to drive safely.
The SH can be divided into segments with each segment providing virtual resources to the swarm of smart cars within its range. The shared pool of resources including video cameras, pressure sensors, and speed monitors need to operate synchronously to tackle the tasks in real time without creating backlogs (i.e., a { \em capacity pool} use case for multi-point synchronization). The challenge here is to deal with coalitions that are short-lived (i.e., coalitions created and destroyed as cars move by) with low synchronization overhead. 

\subsubsection{Bridge Health Monitoring}

This is an example of the~\textit{data capture synchrony} use case.
Strain measurement at bridge joints and other important points in the structure is required to maintain a close watch on the health of a bridge structure. To make high quality measurements, it is necessary to coordinate the data capture operations such that 
they are made when the loading is at a particular configuration. The loading configuration would be measured by the position of the vehicles on the bridge at that instant and their weight. 

The most accurate way of doing such a measurement is to actuate all involved devices (sensors and vehicles) to run the measurement function at the same time instant. If different devices run the measurement function at different time points, a complex reconstruction procedure need to be executed to determine the concurrent loading. 
The fine-grained measurements are only of interest while the vehicles are on the bridge. That is, they do not need to continue to take fine grained position measurements and report them when they are not on the bridge.

\section{System Architecture}\label{sec3}

\begin{figure}[b]
\centering
\includegraphics[height=2.2in]{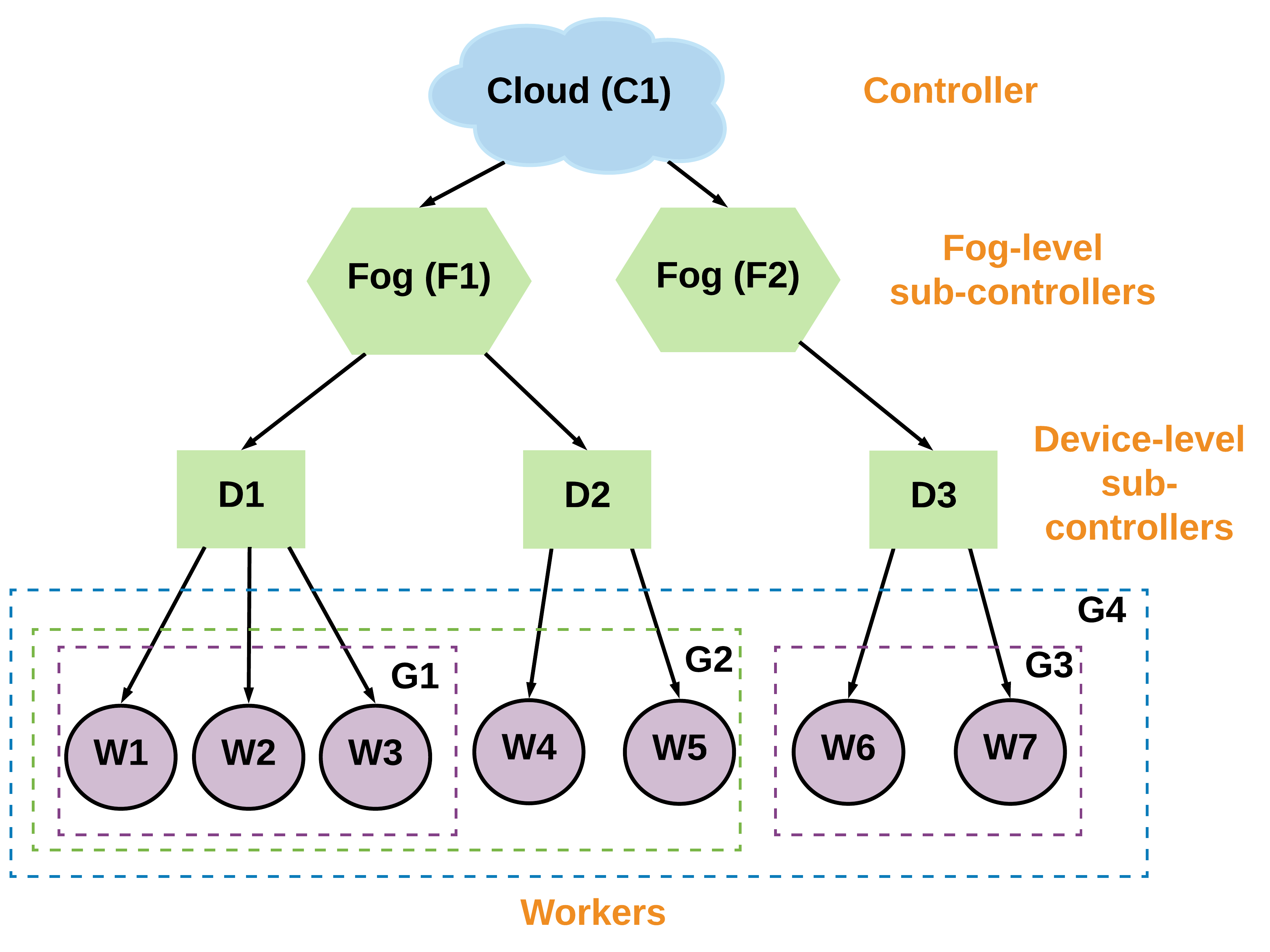}
\caption{Multi-level hierarchical node model showing controller, sub-controllers and worker nodes.}
\label{model}
\end{figure}

In the system, nodes are organized into a multi-level controller-worker tree as shown in  Fig.~\ref{model}  with the cloud at the topmost level of the tree and worker nodes at the leaves of the tree. Workers could be fixed or mobile. Fogs are located between the cloud and workers. The cloud controller has a global view of the entire system while the fog-level sub-controllers have limited view but are located closer to the workers. This architecture is suitable for achieving synchronization in IoT because it permits worker nodes to join, leave or move around, thus, changing membership one sub-controller to the other. The fog-level controllers provide localized and low latency services to the worker which is necessary for getting synchronization in IoT.

Multiple levels of sub-controllers could exist between the cloud and the workers, the sub-controllers could be at the fog and device level.
Tight clock synchronization is assumed across all the levels in the tree by leveraging the hierarchical architecture and the recent advancements in clock synchronization schemes~\cite{ademaj2007time}. Our controller-worker model is inspired by systems such as multi-robot systems~\cite{gautam2012review}, software defined networking (SDN)~\cite{kreutz2015software} and fog computing systems.   

\subsection{Node Model}

Both the controllers and workers in our system follow a single-threaded execution model (i.e., they can only run a single task at a time). Worker nodes are isolated from each other and thus, no direct link exists between them. Workers only communicate with the sub-controller or controller to which they are connected. Communication between a worker and controller is bi-directional, that is data can flow both ways. The scope of the task being run determines what level of controller the workers communicate with. A global scope will require communication to exist between the main controller (cloud-level controller) and workers either directly or through sub-controllers. A non-global scope involves workers communicating with the fog-level controller while a much more localized scope involves workers communicating and working with only the device-level controller. 
 
A heterogeneous system is assumed where worker nodes have varying processing and computational capabilities, thus, worker nodes are expected to have different execution times for the same task. Due to network disruptions, mobility or node failures, disconnections can occur between worker nodes and controllers. Disconnected workers can rejoin the system by connecting to a controller and new workers can only join the system if they connect to a controller.

\subsection{Application and Task Model}

An application written for this model consists of remote and local function calls. The remote function calls invoked by the controllers on sub-controllers and workers are downcalls while calls by the workers on controllers at any level (i.e., one or more of the cloud, fog, device levels) are upcalls. For instance, the root controller can downcall to the fog-level or device-level.

\begin{figure}[t]
\centering
\includegraphics[height=1.7in]{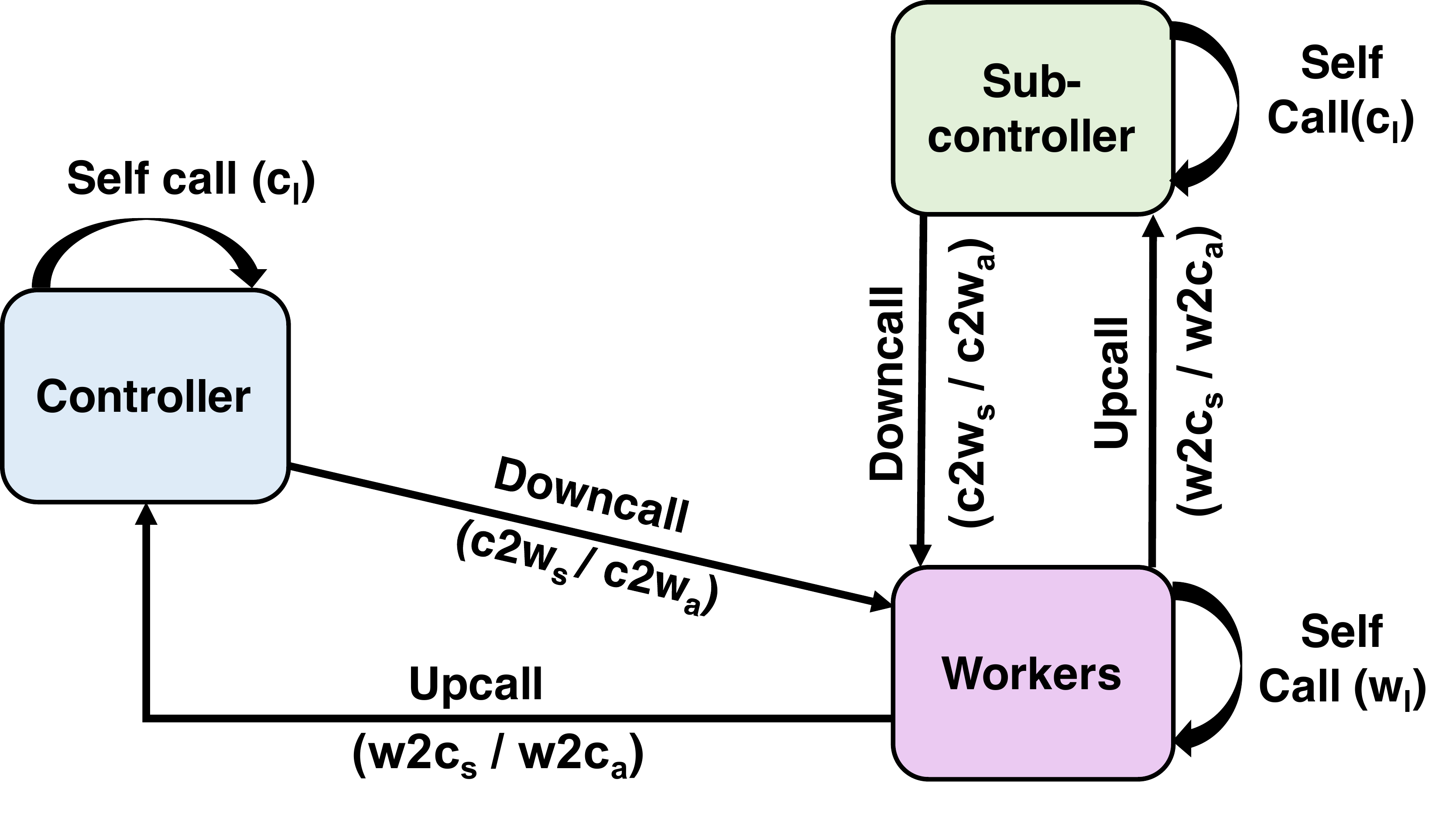}
\caption{Task model showing synchronous, asynchronous and local calls between controllers and workers.}
\label{task}
\end{figure}

The upcalls and downcalls can either be synchronous or asynchronous as shown in Fig.~\ref{task}. A synchronous call blocks the calling node until the remote execution completes while an asynchronous call is non-blocking. Because the upward and downward calls can involve multiple workers, the synchronous execution of the calls is not simple. In particular, the conditions under which the group would provide a valid execution for a synchronous call needs to be defined. 

When a controller makes a synchronous downcall, workers individually sign up for executing the function. Once the controller receives commitments from sufficient number of workers to run the function, the controller will proceed with running the function on the workers. The requirement is to start the execution precisely (assuming the underlying clocks are synchronized) at the same time across all the workers. This is a hard problem because even the signed-up workers can become available to execute the function at different times. Therefore, the function execution needs to be scheduled across the workers such that the start time skew and idle times at the workers are simultaneously minimized.

Tasks are classified based on function calls into the following categories.
\begin{enumerate}
\item \textbf{Controller-to-worker asynchronous call}: This is a call from a controller to its worker nodes to run a task without having strict timing conditions, that is, the controller does not wait during this period. This task is denoted as $c2w_{a}$.
\item \textbf{Controller-to-worker synchronous call}: The controller sends a command to all its worker nodes to start executing a task at the same point in time. The controller waits until all worker nodes have finished executing the task and returned an output. This task is denoted as $c2w_s$.
\item \textbf{Worker-to-controller asynchronous call}: This is a call from a worker node to one of its controllers to run a task. The worker does not wait during this period and continues with its own execution plan. This task is denoted as $w2c_a$
\item \textbf{Worker-to-controller synchronous call}: The worker node waits for an acknowledgement from the controller that it has finished executing the task associated with the call. This task is denoted as $w2c_s$ 
\item \textbf{Self calls}: This is a call from a node to itself. It could be either a worker node calling itself or a controller calling itself. A self triggered worker task is called a local worker task and is denoted as $w_l$ while a self triggered controller task is denoted as $c_l$.
\end{enumerate}

\section{Algorithms}\label{sec4}

Two dynamic synchronous scheduling algorithms with redundancy built into them for scheduling tasks with varying time requirements onto a bunch of IoT nodes are developed. The unique aspects of the algorithms are an adapted publish-subscribe status update scheme, quorum checking, redundancy and the local scheduler which are explained in the following subsections. The main focus of the algorithms is to ensure that synchronous tasks are scheduled to run at the same time across all workers. Tasks are gotten from the application graph and arrive in order of precedence such that a task depending on output data from a previous task cannot be spawned until the previous task has finished executing. 

One input to the synchronous scheduling algorithms is a set of available tasks. The algorithms are expected to output a schedule of tasks that minimizes the overall execution time and ensures that the desired level of synchronization is met with respect to the quorum requirements. In addition to the notations introduced in Section~\ref{sec3}, the following notations are used in the algorithms. A description of symbols used in this paper is shown in Table~\ref{proj_table}.

\begin{table}[!ht]
\renewcommand{\arraystretch}{1.2}
\caption{Algorithm notations.}
\label{proj_table}
\centering
\begin{tabular}{l|l}
\bfseries Symbol & \bfseries Description \\ \hline
$T_{curr}$  &  current task to be scheduled \\ 
\hline
$T_{rbq}$  &  ratio-based quorum check task \\  
\hline 
$T_{cbq}$  &  cluster-based quorum check task \\ 
\hline 
$T_{update}$  &  status update task \\
\hline 
$t_{avail}$   & available time of a worker \\ 
\hline 
$\lambda$  & delay before re-attempting quorum \\ 
\hline 
$\delta$  & predicted quorum check time \\  

\end{tabular}
\end{table}

\subsection{Status Update Scheme}

Workers need to update the controller of their availability to partake in synchronization upon reaching a synchronization point. The controller processes this information serially due to its single-threaded nature. Thus, the message complexity of the update sending process scales up linearly with the number of workers; thus, increasing the time required to achieve synchronization significantly. 

To mitigate this problem, the popular publish-subscribe scheme is adapted to reduce the number of update messages sent to the controller. Worker nodes are grouped into clusters and each cluster is assigned a local broker, where all the workers publish their availabilities. They also subscribe to the broker to know the peer availabilities and when a worker detects that the required number of peer workers are present, it will publish a group availability message at the broker. The controller subscribes to the group availability message but not the not local availabilities. So, it ends up processing far less with the broker. 

\subsection{Quorum Types}

Quorum checking is done by workers to probe the controller on whether the required conditions for proceeding to run the synchronization task are met, thus providing some $QoS_{ync}$. Two types of quorum conditions are considered. One based on ratio of workers available and the other on cluster representation. A worker is said to be available if it has finished executing its previous task and is physically present to run the next task. The two types of quorum checking are ratio-based (launched on workers by running $T_{rbq}$) and cluster-based (launched on workers by running  $T_{cbq}$) quorums, both of which are presented in the taxonomy in Section~\ref{qcheck}.

\subsection{Synchronous Scheduling Scheme}

\begin{figure}[bt]
\centering
\includegraphics[width=3.3in]{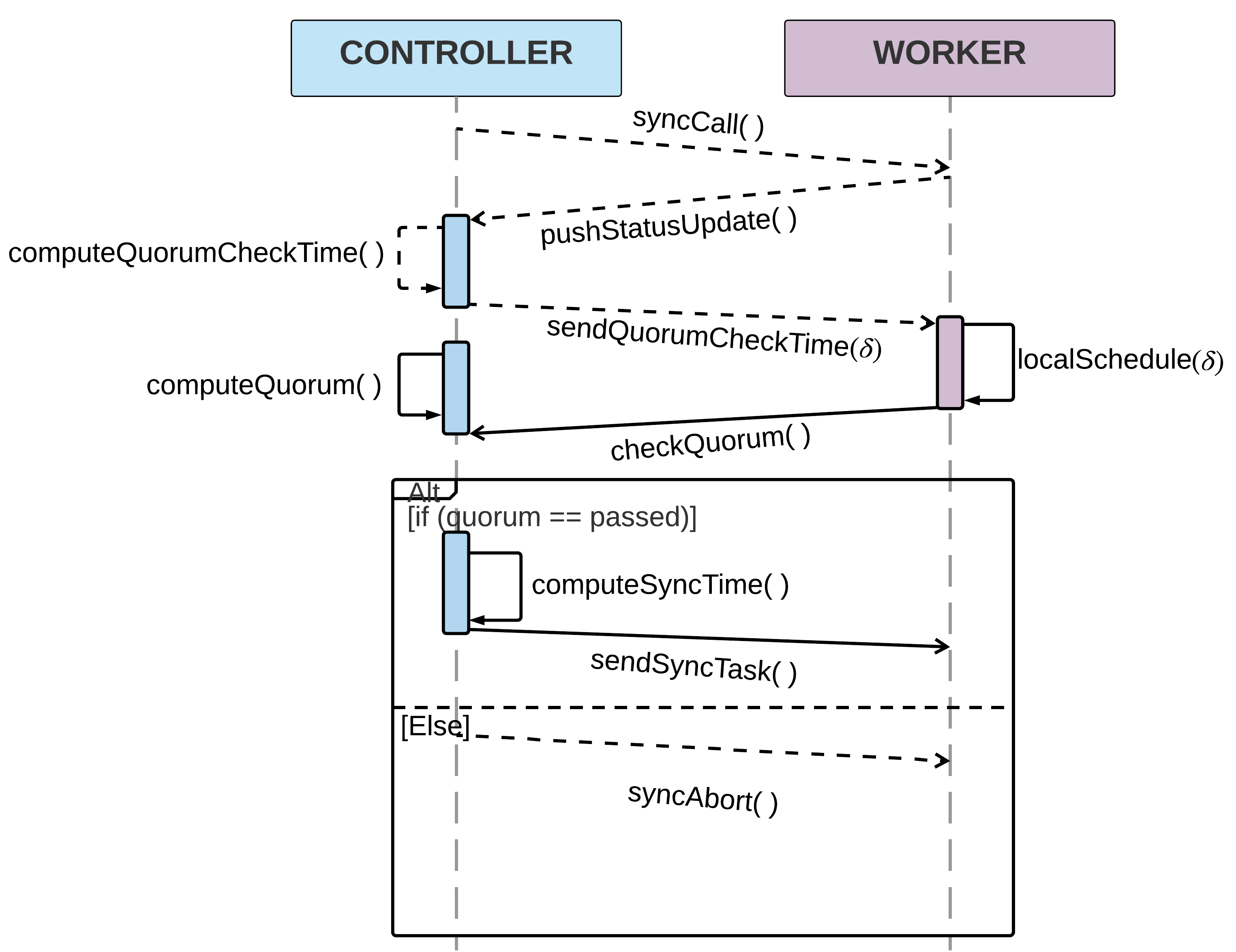}
\caption{Sequence diagram showing the interactions between the controller and workers at synchronization point.}
\label{sequence}
\end{figure}

As seen in Fig.~\ref{sequence}, the controller sends a sync call ({\tt syncCall()}) to the workers on getting to a synchronization point. The workers update the controller of their status ({\tt pushStatusUpdate()}) either using a naive scheme where all workers send update messages to the controller or using the adapted publish-subscribe update scheme. The pseudocode for the algorithm is shown in Algorithm~\ref{pseudo}. 

The controller computes the predicted quorum check time $\delta$ based on the status update from workers and sends the time to local schedulers on workers ({\tt sendQuorumCheckTime($\delta$)}). The local scheduler on a worker tries to minimize wait time by comparing the available time of the worker with $\delta$ and estimates if a local task (provided there is one in the local task queue) can fit within the gap as shown on~\textit{Line 6} of Algorithm~\ref{pseudo}. The local scheduler ({\tt localSchedule($\delta)$}) is an optimization scheme to minimize the idle time on workers that become available earlier than other workers as shown in Algorithm~\ref{local}. Asynchronous and local worker tasks are scheduled as they become the current task at the earliest available times on workers.

\begin{algorithm}[tbp]
\SetAlgoLined

Let $A$ = {set of available tasks}

\textit{SyncSchd(A)}:

\tab \textbf{while} $T_{curr} \neq \phi$ \textbf{do}:
 
\stab \textbf{if}  \textit{type($T_{curr}$) = $c2w_s$}:
 
\mtab \textit{pushStatusUpdate()}

\mtab $\delta$ = \textit{computeQuorumCheckTime()}
    	
\mtab \textit{localSchedule($\delta$)}

\mtab \textit{revise\_available\_times($t_{avail}$, max\_t\_avail)}

\mtab \textit{schedule(quorum check task)}

\textbf{\mtab Time-based redundancy:}

\mtab r = \textit{getAvailableRatio()}

\mtab \textbf{if} \textit{r $>=$ sync degree}:

\ltab \textit{schedule($T_{curr}$)}

\mtab \textbf{else if} \textit{r $<$ sync degree} \&\& retries available:

\ltab \textit{SyncSchd(A, $t_{avail} + \lambda$)}

\mtab \textbf{else if} \textit{r $<$ sync degree} \&\& no more retries:

\ltab \textit{syncAbort()}

\textbf{\mtab Component-based redundancy:}

\mtab e = \textit{computeRedundancy()}

\mtab \textbf{if} \textit{e $>=$ required redundancy}:

\ltab \textit{schedule($T_{curr}$)}

\mtab \textbf{else}:

\ltab \textit{syncAbort()}

\mtab \textit{remove($A, T_{curr}$)}
 
\stab  \textbf{else if} \textit{type($T_{curr}$) = $c2w_a$} $\parallel$ \textit{type($T_{curr}$) = $W_l$}:

\mtab \textit{revise\_available\_times($t_{avail}$, get\_ctrl\_avail\_time())}

\mtab  \textit{schedule($T_{curr}$)}

\stab \textit{remove($A, T_{curr}$)}

\caption{Pseudocode for synchronous scheduling algorithms.}
\label{pseudo}
\end{algorithm}

The two types of the dynamic synchronous scheduling algorithm are the time-based redundancy where synchronization is attempted based on waiting and a capped number of retries, and the component-based redundancy where at least a certain number of workers within a group are expected to be present before quorum can be passed.

\subsubsection{Synchronous Scheduling Algorithm with Time-Based Redundancy}\label{timeAlgo}

Here, the synchronization degree represents the ratio of workers that must be available before the synchronous task can be run. Whenever the ratio-based quorum check task $T_{rbq}$ is scheduled (\textit{Line 9}), the controller computes the ratio of available workers. If the ratio of available workers is greater than or equal to the expected synchronization degree~(\textit{Line 12}), the controller computes the expected synchronous task start time and the synchronous task is scheduled. 

If the synchronization degree is not met and there are more retries left (\textit{Line 14}), execution is delayed for time $\lambda$ and the process is retried. However, if the synchronization degree is not met and there are no more retries left, the synchronous task is failed (\textit{Line 17}). An example where time-based redundancy is useful is in bridge health monitoring. If while trying to take measurements a failure occurs, the application can proceed with its processing and repeat the bridge strain measurement at a later point in time. 

\subsubsection{Synchronous Scheduling Algorithm with Component-Based Redundancy}

Here, workers are always part of a logical cluster as they move around in the system (e.g., vehicles). At a synchronization point, after the update of the workers' status at the controller (\textit{Line 5}) and the triggering of the local schedulers (\textit{Line 7}), the workers run the quorum check task ($T_{cbq}$). The controller computes and checks whether the required level of redundancy is met by each cluster (\textit{Line 19-20}). This computation is done by comparing the number of workers available in each cluster with the expected number of devices per cluster. In the event of a successful quorum, the synchronous task is scheduled to run. If the desired level of redundancy is not met, the synchronous task is failed (\textit{Line 23}).

In this algorithm, a synchronous task is marked as successfully completed if and only if at least the required number of workers per cluster returns a result after executing the synchronous task. Thus, the synchronization result is abstracted at the cluster level. If the desired redundancy in output is not met, the synchronous task is considered to have failed. An example of where component-based redundancy is useful is in the drone transportation scheme. The drones are expected not to perform the transportation task if the desired number of backup drones are not available. 

\subsection{Local Scheduler}

The local scheduling is an optimization scheme developed to minimize the waiting of workers that get to the synchronization point earlier than other workers. Since the workers do not communicate directly with each other but only with the controller, workers on getting to a synchronization point have no information about other workers. So, by running a local scheduler, a worker compares its current time with the predicted finish time $\delta$ across all workers as shown in Algorithm~\ref{local}. 

The local scheduler would run a local task if the gap between the predicted availability of the other workers $\delta$ and the local node's availability $t_{avail}$ is greater than the local task's length. Otherwise, the local scheduler would not schedule any local task and let the worker sit idle.

\begin{algorithm}[htbp]
\SetAlgoLined

\textit{localSchedule($\delta$)}:

\textit{L} = \{local worker task queue\}

\textbf{while} \textit{get\_tasks(L) =} $T_{local}$ and $T_{local} \neq \phi$ \textbf{do}:

\tab \textbf{if} $t_{avail}$ + $t_l \leq \delta$:

\stab \textit{schedule($T_{local}$)}

\stab \textit{revise\_available\_times($t_{avail}$, max\_t\_avail)}
    
\tab \textbf{else}:

\stab continue
    
\caption{Local scheduling algorithm for minimizing idle time}
\label{local}
\end{algorithm}

\section{Experiments and Results}\label{sec5}

First, experiments are performed to measure the impact of controller location (i.e., fog or cloud) on synchronization. Then, further experiments are conducted to measure specific attributes of the performance of the proposed synchronization algorithms. Finally, the performance of the proposed synchronization algorithms are evaluated by comparing them with barrier synchronization~\cite{khayyat2013mizan, jakovits2012stratus} and time slotted synchronization~\cite{vogli2015fast}.

\subsection{Impact of Controller Location on Synchronization}

In this experiment, traces from the OpenCloud Hadoop cluster from Carnegie Mellon University Parallel Data Lab \footnote{http://ftp.pdl.cmu.edu/pub/datasets/hla/dataset.html} are used. The workload is split into short tasks ranging from 0.5s to 4s and long tasks ranging from 5s to 12s. The controller-worker delay is varied from 5ms to 500ms. Performance is measured using synchronization rate (SR) which is the number of synchronizations per unit time.

\begin{figure}[t]
\centering
    \includegraphics[width=0.7\linewidth]{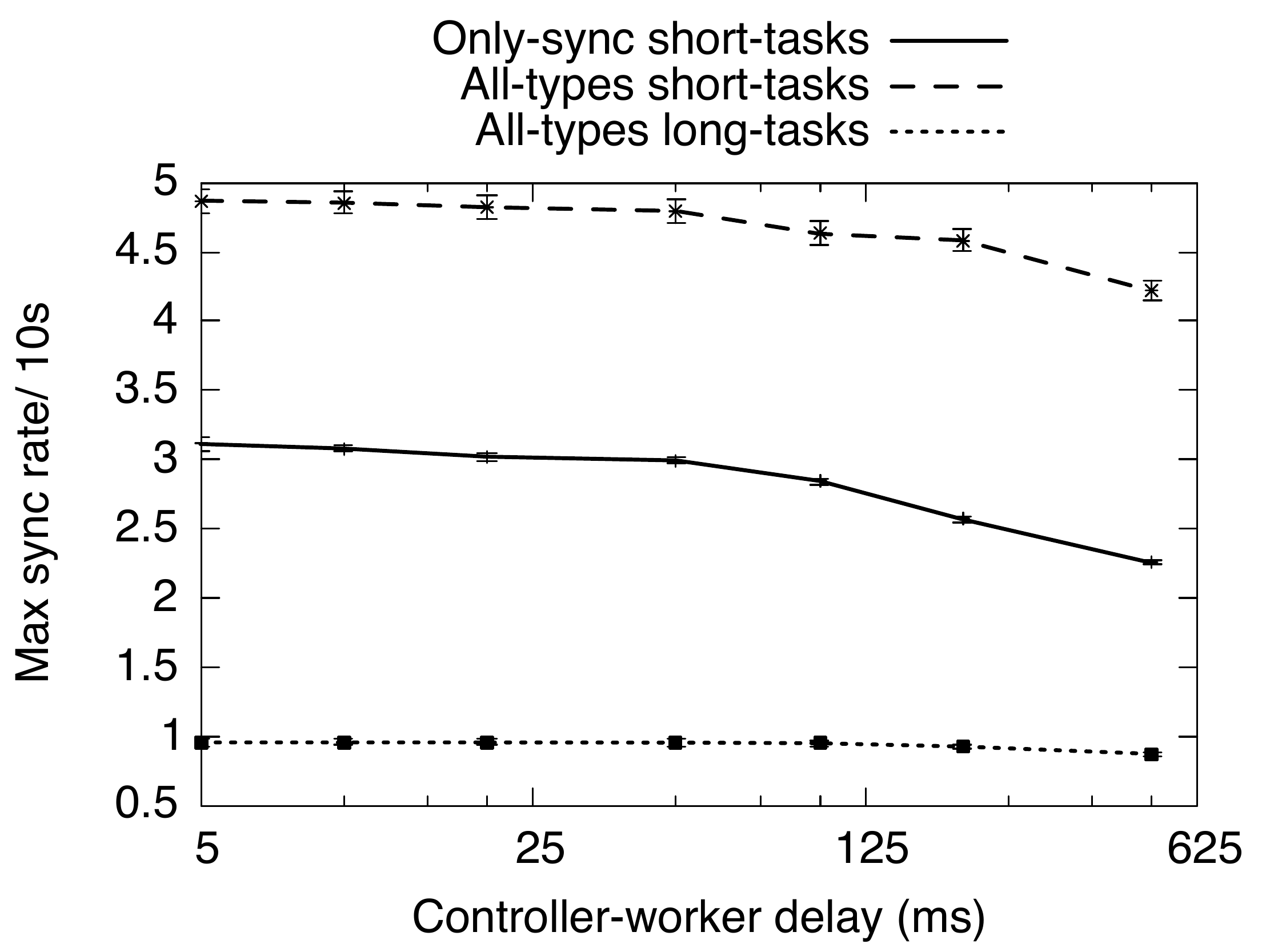}\par\caption{Maximum synchronization rate per 10s for varying controller-worker delays.}\label{edgecloud}
\end{figure}

From Fig.~\ref{edgecloud}, it can be observed that a task graph consisting of short jobs has much higher SR compared to a task graph consisting of long tasks. Also, increasing the controller-worker delay from 5ms to 500ms has very little on impact on the task graph with long tasks compared to short tasks because the long tasks take a significant portion of the overall runtime thus minimizing the impact of the controller-worker delay. A task graph with short tasks consisting of synchronous, asynchronous and local tasks has a higher SR compared to one consisting of only synchronous tasks. This is because with only synchronization tasks, status updates and quorum checking needs to be performed at each synchronization point therefore adding more overhead to the overall runtime. Having controllers closer to the workers increases the SR for short running tasks, thus, making the case for fog-resident controllers as opposed to cloud-resident controllers.

\subsection{Configuration of Synchronization Experiments}

The following configurations are used to define the system, applications and environmental parameters. The wide range of parameter variation allows for a higher degree of exploration into many aspects of the system. Measurement traces from experiments using Dropbox between the period of 28th June to 3rd July 2012~\cite{gracia2013actively} are used as the task dataset. The execution times vary with a minimum time of 23$s$ and a maximum time of 269$s$.

To model the mobility of worker nodes, the Shanghai (China) taxi GPS report for Feb 20, 2007~\cite{liu2010towards} is used. The report consists of 4316 taxis reporting their location, speed, angle of movement and occupancy at given intervals over a period of 24 hours. Each taxi has a unique identification number. The taxi traces are pre-processed and only the location and timestamp details are extracted. The taxis upload their details at irregular intervals in trace, varying from 15s to 63s. Re-sampling is done at 30s and the position of the taxis are gotten at regular intervals, thus having a total of 2880 time points.


The parameters used in the simulations are as follows. (i) \textit{Synchronization degree}: ratio of the total machines required to pass quorum. (ii) \textit{Minimum cluster size}: the minimum number of workers that must be present in a cluster before it can be formed. (iii) \textit{Wait time}: The amount of time that should elapse before attempting quorum checking again. (iv) \textit{Quorum retries}: the maximum number of times quorum checking is allowed. (v) \textit{Worker size}: The maximum number of workers that can be present in the system at any point in time. (vi) \textit{Number of clusters}:  The maximum number of clusters that can be formed at any point in time. (vii) \textit{Prediction accuracy}: A measure of how accurately the predictor predicts the finish time of tasks across all workers prior to the synchronization point. 

\subsection{Default Parameter Values and Measurements}

The following parameters are fixed in the simulations unless otherwise stated. The number of independent runs of each simulation is $100$ while each task graph is continuously run in each simulation for $200$ times, the communication cost between machines is set to $200ms$, status update cost is set to $1s$, synchronization degree is set to $0.7$ and $\lambda$ set to $20s$. Workers randomly fail with a probability of $0.1$ after each task. The probability of a new machine joining is set at $0.1$, but machines can only join at the start of the execution of a new run of the task graph. This is done to ensure that joining machines will have all the necessary data required to run all tasks down the task graph.

Task graphs consist of $30$ tasks in total with varying number of synchronous tasks. $10$ task graphs are used to represent different applications in our simulation runs. Heterogeneity among the worker nodes is introduced by making the execution time of a task on multiple workers follow a Gaussian distribution.


The following parameters are measured in the simulations. (i) \textit{Runtime}: the time taken for a single run of a task graph. It is gotten by dividing the total execution time of a simulation by the number of runs. In our results, the execution time is normalized by the number of sync points. (ii) \textit{Extra quorum attempts}: the total number of times the quorum check process was attempted after the initial attempt at all synchronization points. (iii) \textit{Failed sync tasks}: The total number of sync tasks that failed after exceeding the total number of quorum retries or due to incomplete results from clusters.

\subsection{Experiments, Results and Discussions}

\subsubsection{Scalability of the adapted publish-subscribe update scheme}

\begin{figure}[b]
\centering
    \includegraphics[width=0.7\linewidth]{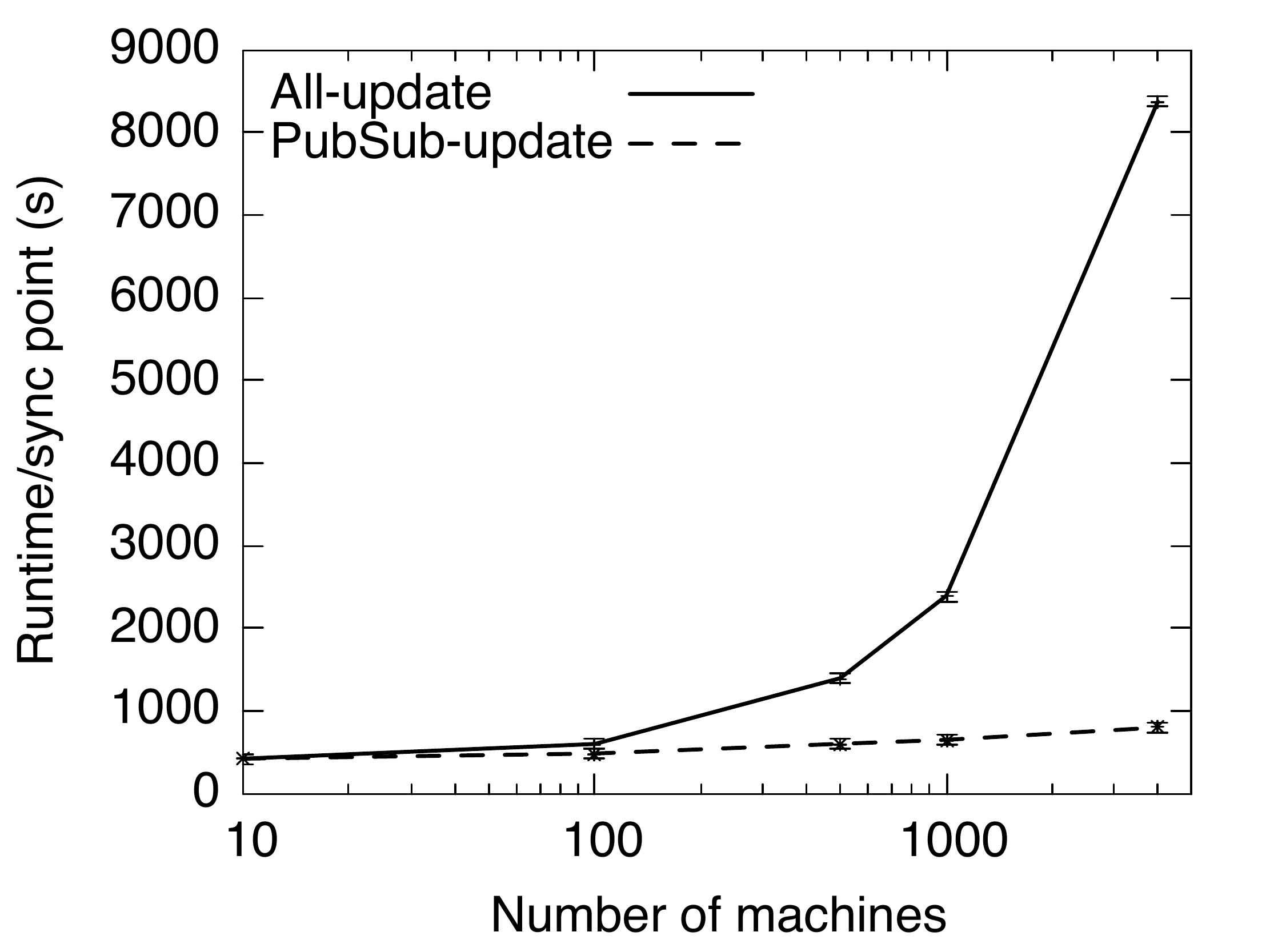}\par\caption{Runtime per sync point comparing all-worker update sending and the publish-subscribe update scheme for time-based redundancy.}\label{pbtime}
 \end{figure}
 
 \begin{figure*}[t]
\begin{multicols}{3}
    \includegraphics[width=\linewidth]{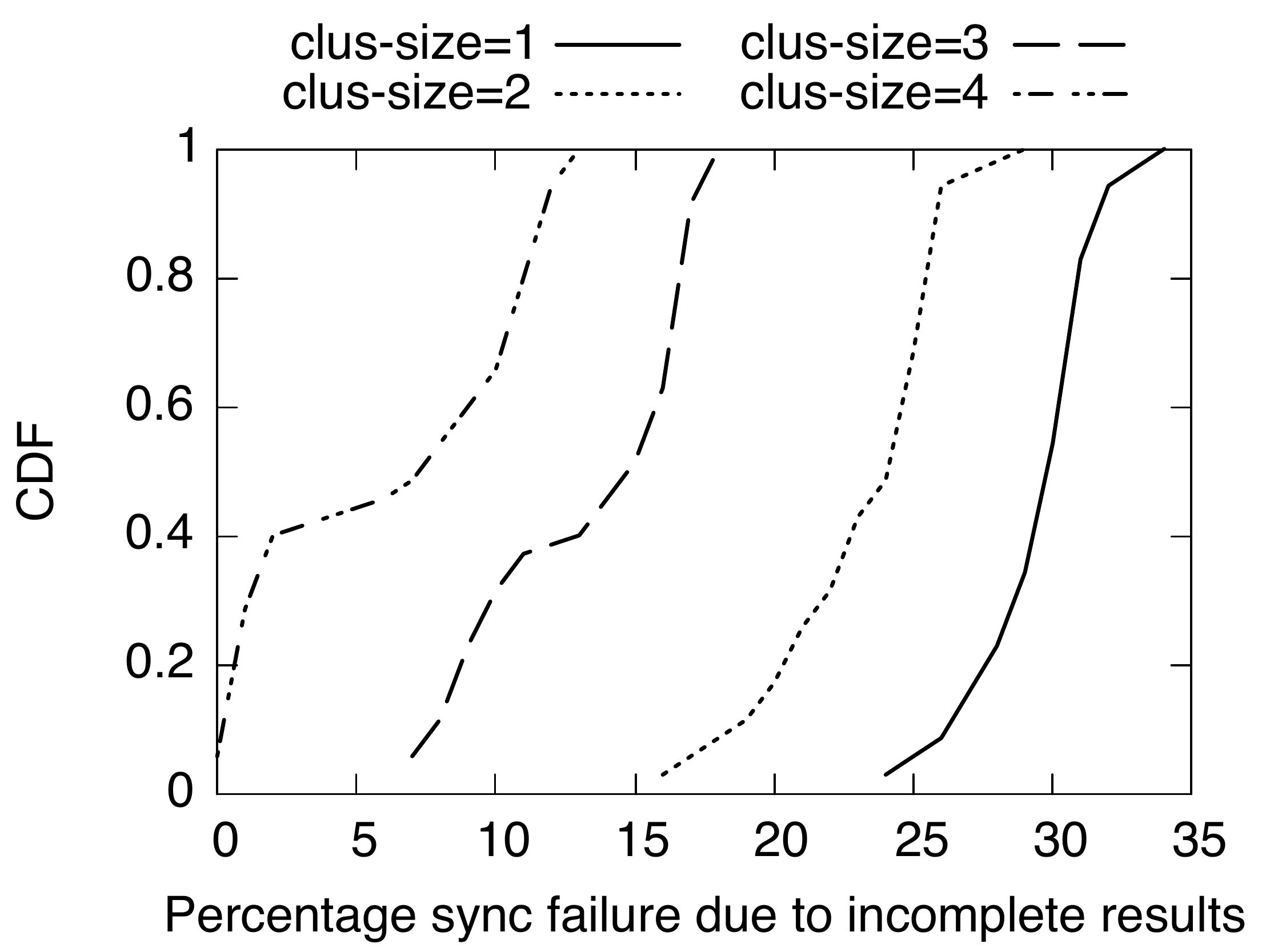}\par\caption{CDF showing percentage of sync task failures due to incomplete results from clusters for minimum cluster sizes ranging from $1$ to $4$ for component-based redundancy.}\label{redunincomplete}
    \includegraphics[width=\linewidth]{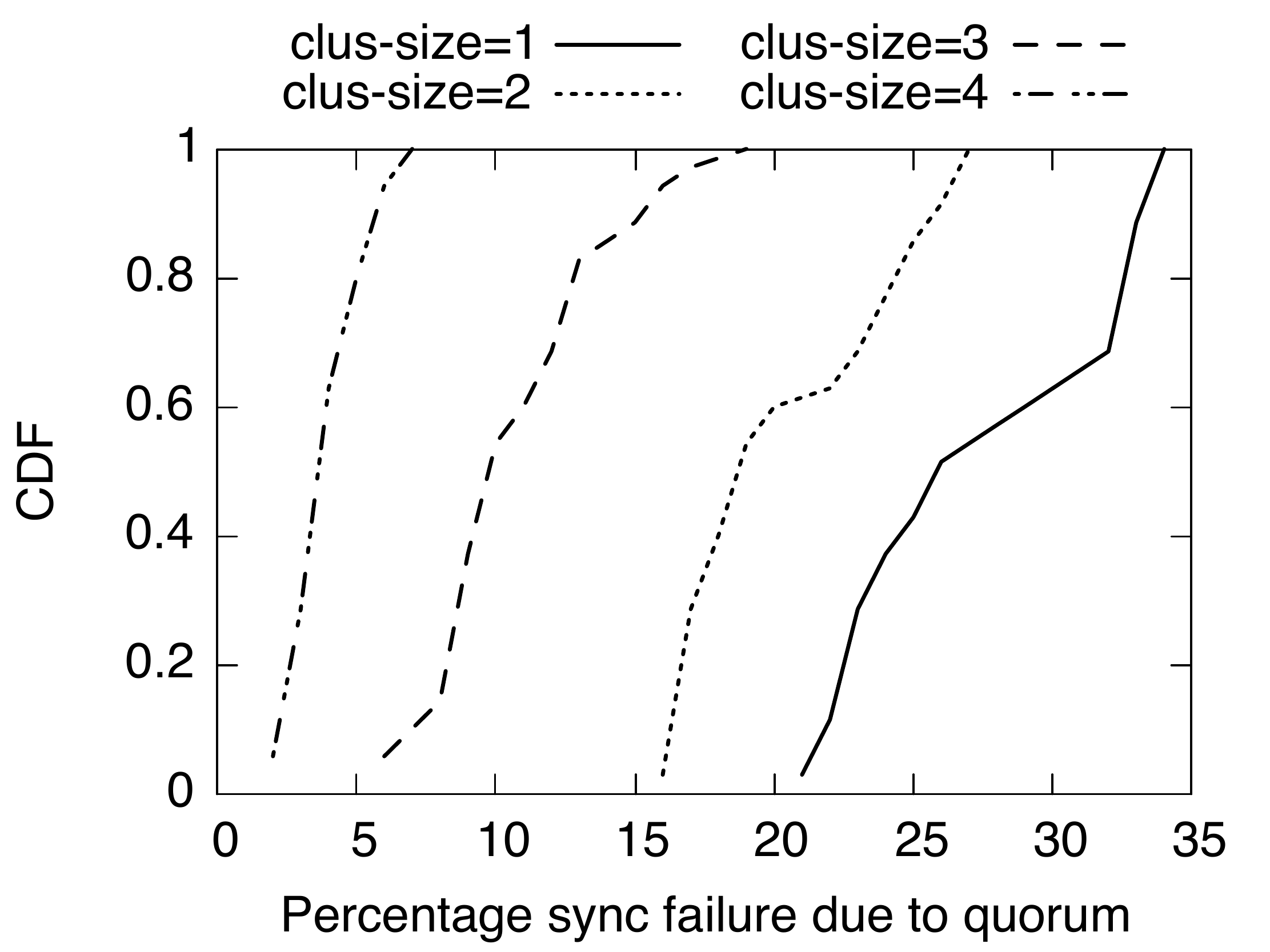}\par\caption{CDF for the percentage of sync task failures caused by failed quorum for different minimum cluster sizes for component-based redundancy.}\label{redunquorum}
    \includegraphics[width=\linewidth]{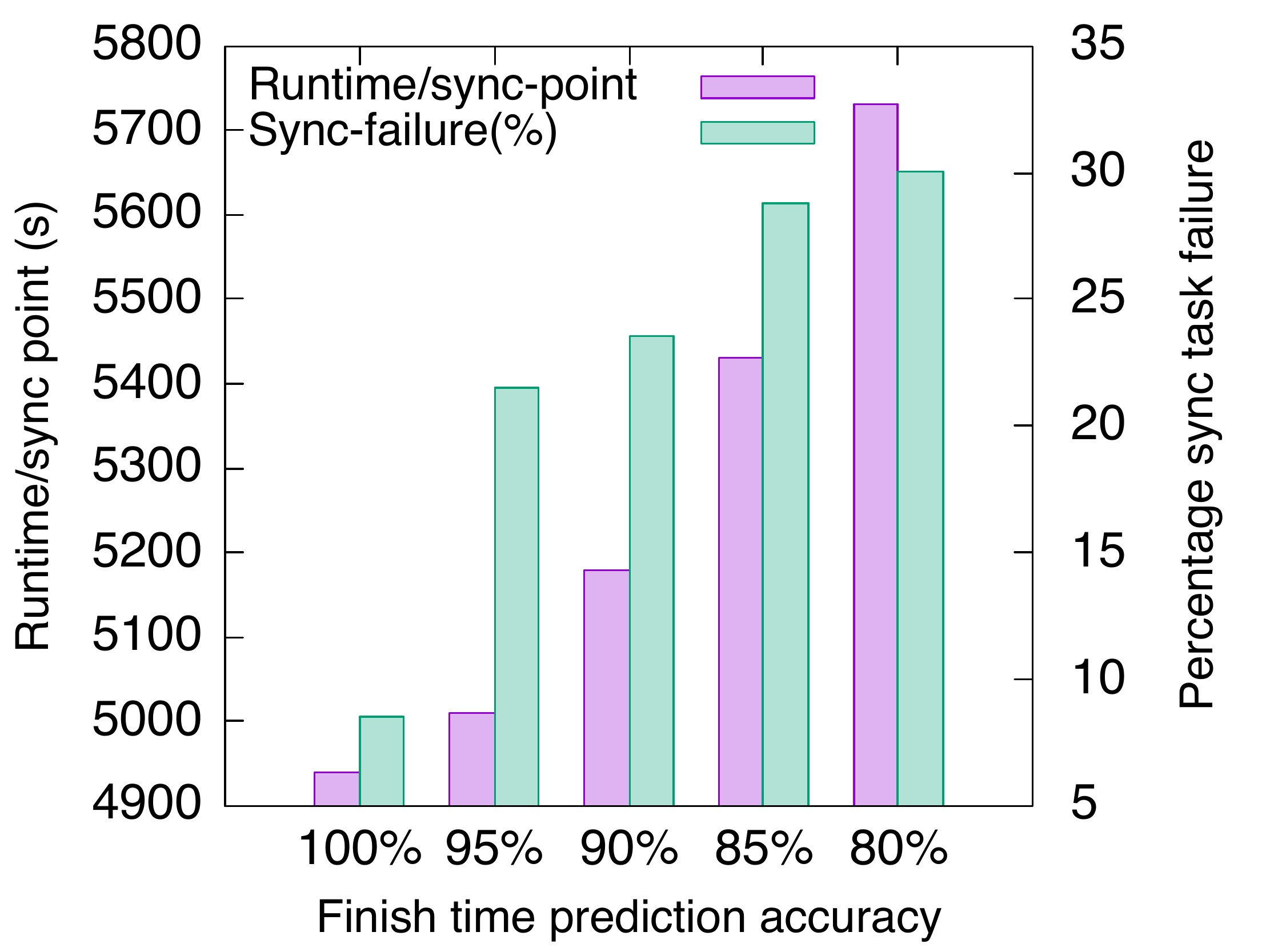}\par\caption{Runtime per synchronization and percentage synchronization task failure for task finish time prediction accuracy varying from $80$\% to $100$\%.}\label{accuracyTF}
\end{multicols}
\end{figure*}

On getting to a synchronization point, workers update 
their associated controller of their availability for running synchronous tasks. The updates are serially processed by the controllers due to their 
single-threaded nature. 
This can pose a serious bottleneck problem as the number of workers increase. 
To alleviate this, a publish-subscribe mechanism is adapted. The benefits of
the mechanism for are shown in Fig.~\ref{pbtime} while varying the number of workers from $10$ to $4000$.
 
Fig.~\ref{pbtime} shows the runtime per synchronization point for the publish-subscribe and all-worker update methods for the ratio-based quorum checking. From the graph, it can be observed that as the number of workers increase the runtime per synchronization point increases at a similar rate with respect to the number of workers for the all-worker update while for the publish-subscribe update there is no significant increase in the runtime per synchronization point as the number of workers increase. This is because regardless of the number of workers in the system, only a constant number of update messages is being sent to the controller in the publish-subscribe update method.

 \subsubsection{Component redundancy}
 
In component-based redundancy, the workers are grouped into clusters. To reach a quorum to execute a synchronous task, at least a given number of workers must be available in each cluster. The synchronization task is considered successful if at least one worker from the number of workers in the cluster completes the execution of the task and returns the expected output to the controller, otherwise, the synchronous task is failed.
 
The minimum required number of worker(s) per cluster is varied from $1$ to $4$, the synchronization task failures are measured and shown in Figs.~\ref{redunincomplete} -~\ref{redunquorum}. Figs.~\ref{redunincomplete} and~\ref{redunquorum} show the synchronization tasks failure percentage due to failed quorum and incomplete results from clusters, respectively. The percentage of synchronization task failure due to failed quorum and incomplete results from clusters both decrease as the minimum cluster size increases. This is due to the fact that the probability of having workers show up during quorum increases as the minimum cluster size increases.

\subsubsection{Impact of finish time prediction accuracy}

The local scheduler uses the predicted availability of the workers to determine whether it can schedule a local task before starting the synchronous task. In Fig.~\ref{accuracyTF}, the impact of finish time prediction accuracy on the runtime per synchronization point and the synchronization task failure rate with component-based redundancy and publish-subscribe update schemes are measured.

When the prediction accuracy is $100$\%, it was observed that the smallest runtime per synchronization is $4940s$ and the smallest synchronous task failure rate of $8$\%. While at a finish time prediction accuracy of $80$\%, an average of $5730s$ was observed for the runtime per synchronization point and a $30$\% synchronization task failure. The runtime per synchronization point increases by $16$\% while the percentage synchronization task failure increases by $263.3$\% as the finish time prediction accuracy reduces from $100$\% to $80$\%. This shows the high impact that the finish time prediction accuracy has on the success of the synchronization task. 
 
\subsubsection{Performance Evaluation}

\begin{figure}[b]
\centering
      \includegraphics[width=0.7\linewidth]{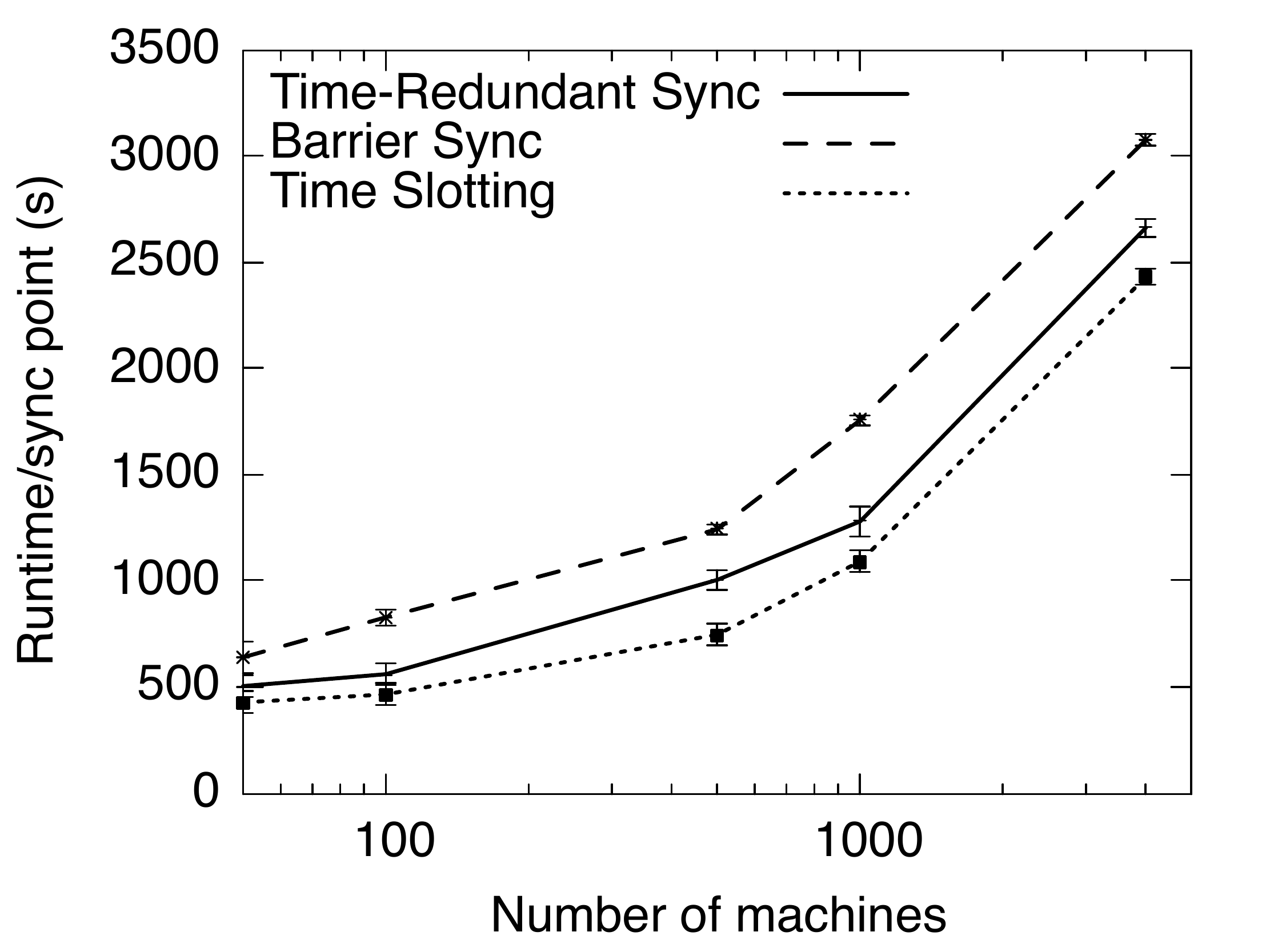}\par\caption{Runtime per sync point comparing the proposed time-redundant synchronization algorithm with barrier and time slotted synchronizations.}\label{timeTime}
 \end{figure}
 
 \begin{figure*}[t]
\begin{multicols}{3}
    \includegraphics[width=\linewidth]{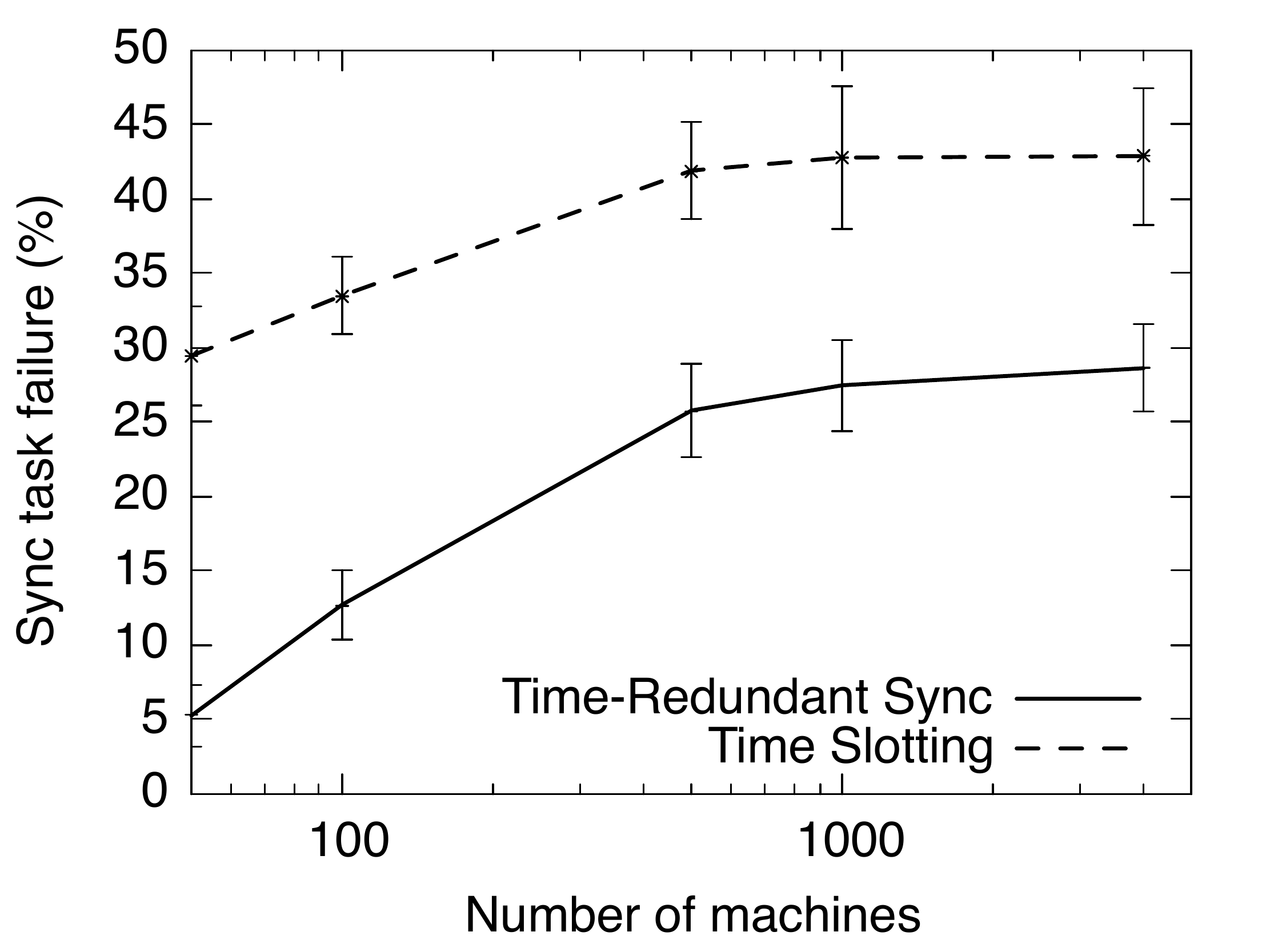}\par\caption{Percentage of sync task failures caused by failed quorum for time-redundant synchronization algorithm vs time slotted synchronization.}\label{timeFail}
    \includegraphics[width=\linewidth]{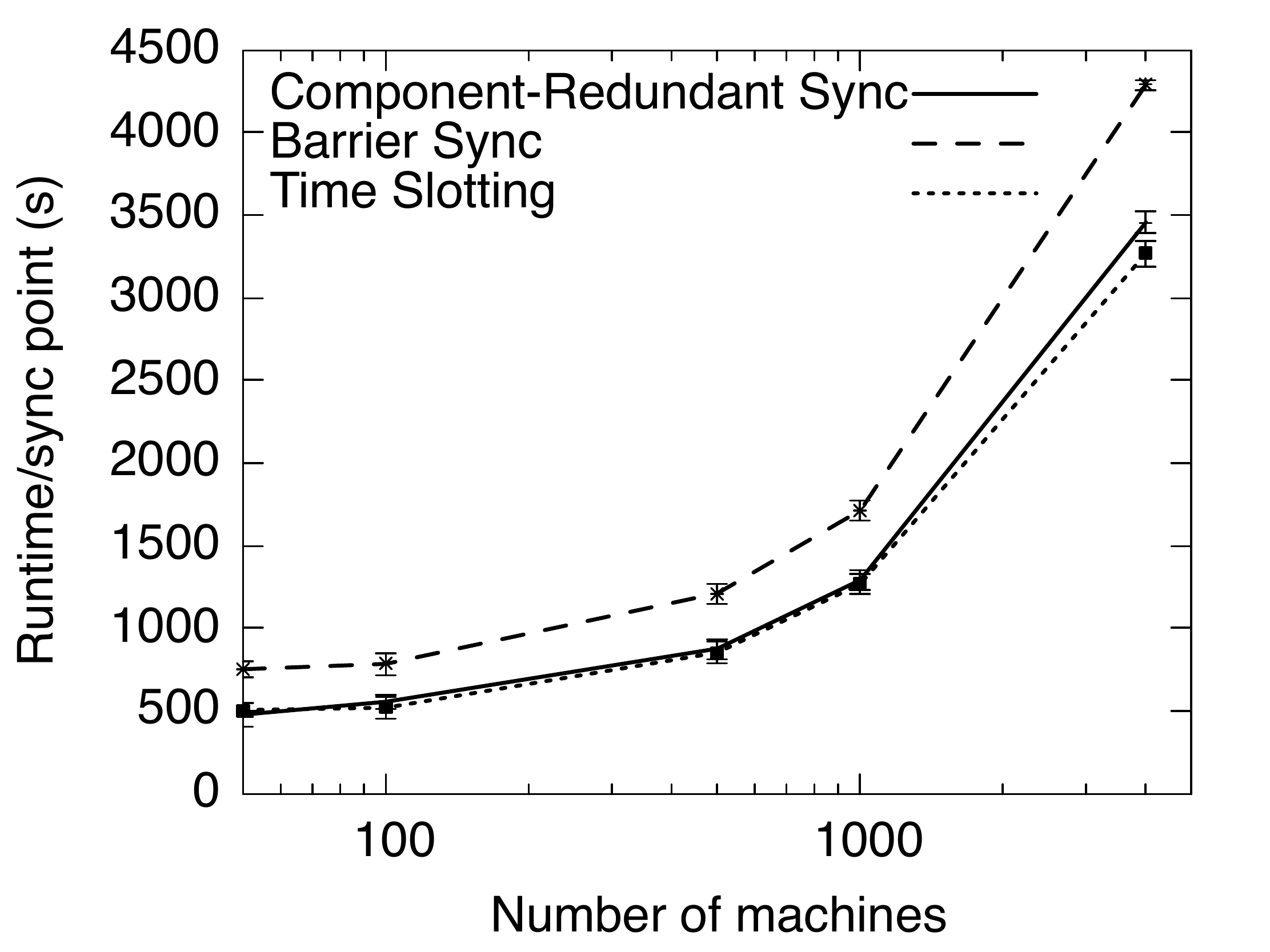}\par\caption{Runtime per sync point comparing the proposed component-redundant synchronization algorithm with barrier and time slotted synchronizations.}\label{componentTime}
    \includegraphics[width=\linewidth]{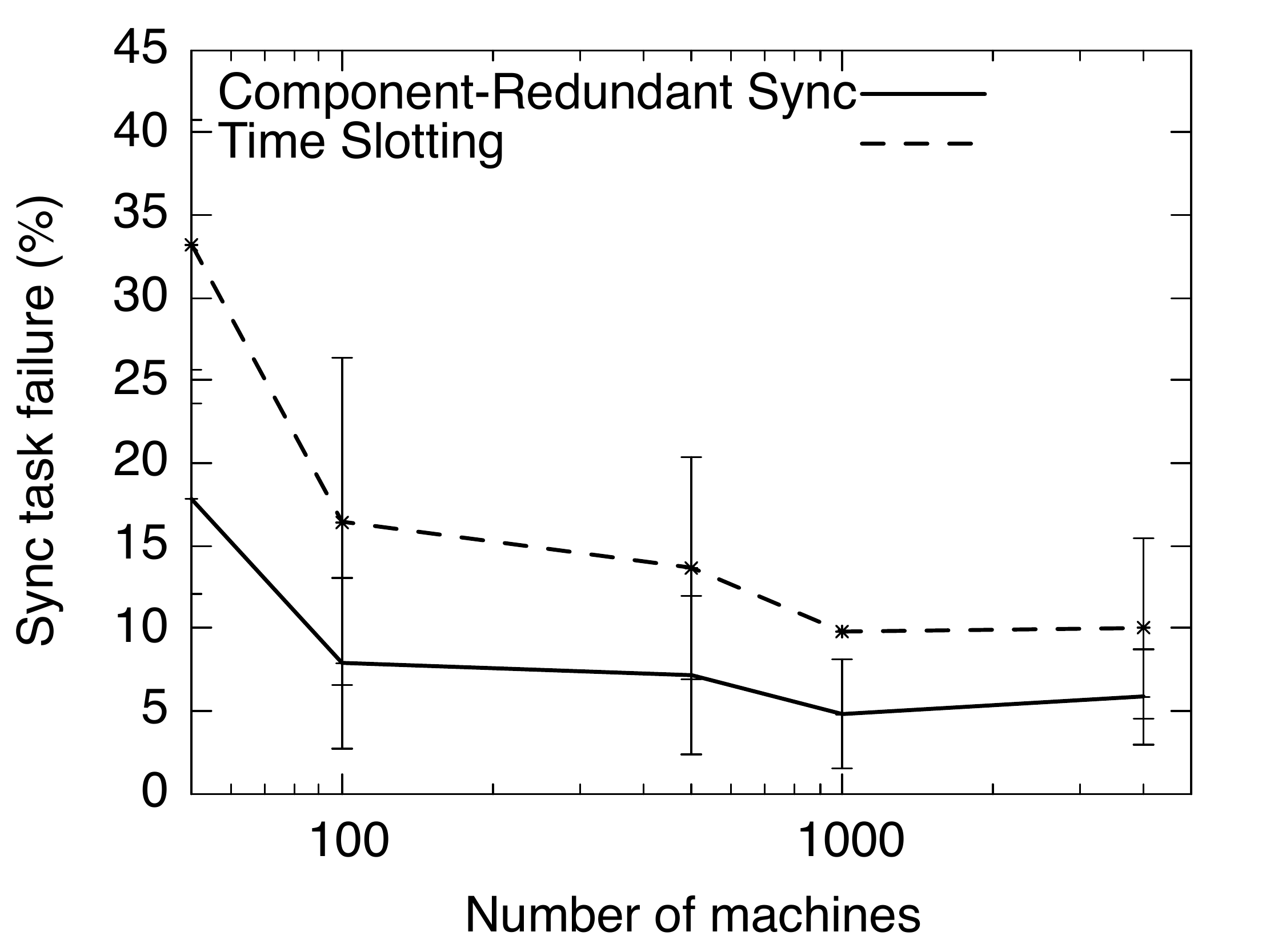}\par\caption{Percentage of sync task failures caused by failed quorum for component-redundant synchronization algorithm vs time slotted synchronization..}\label{componentFail}
\end{multicols}
\end{figure*}
 
The performance of the proposed synchronization schemes  (time-redundant and component-redundant synchronization) is evaluated by comparing them with barrier synchronization and time slotted synchronization. In barrier synchronization, on getting to a sync point, workers send update message to the controller and wait until a signal is received from the controller saying that they can proceed to run the sync task. The condition for proceeding with the barrier execution is that all the workers must reach the sync point. 

In time slotted synchronization, the workers' executions are split into time slots. Dedicated synchronization time slots are chosen with the hope that workers will be available to run the sync task at the specified time slot. The dedicated synchronization time slots are chosen by fixing the slots at $\mu + 1.5\sigma$ (accounts for a 86.6\% accuracy), where $\mu$ is the average execution time and $\sigma$ is the standard deviation.

Figs.~\ref{timeTime} -~\ref{componentFail} show the runtime per synchronization point and percentage synchronization task failure for the synchronous scheduling algorithm with the proposed time-redundant and component-redundant synchronization algorithms, barrier synchronization and time slotted synchronization, respectively. The minimum cluster size for component redundancy is fixed as 3 while comparing the synchronization schemes. It can be observed from Figs.~\ref{timeTime} and~\ref{componentTime} that barrier synchronization has the highest runtime, followed by the time-redundant synchronization algorithm and then time slotted synchronization. Barrier synchronization takes longer because faster workers need to wait for stragglers at the barrier and cannot proceed until the slowest worker reaches the barrier. This is unlike the time-redundant synchronization algorithm that was proposed here, where the synchronization point is moved dynamically depending on the availabilities of workers and also, the local scheduling mechanism is used to minimize wasted work cycles due to waiting. Time slotted synchronization is faster because there are  dedicated synchronization slots that are not moved regardless of workers availabilities.

Figs.~\ref{timeFail} and~\ref{componentFail} show the percentage synchronization task failure due to failed quorum for the proposed time-redundant and component-redundant algorithms, and the time slotted synchronization scheme. The percentage of sync task failures for the time slotted synchronization scheme is higher than that for the proposed time-redundant and component-redundant synchronization algorithms as shown in Figs.~\ref{timeFail} and~\ref{componentFail} respectively. In Fig.~\ref{componentFail}, it can be observed that the percentage of sync task failures reduces as the number of workers increase. This is because there are more devices per cluster and thus, there are more redundant devices which increases the chances of reaching quorum. Time slotted synchronization have higher sync task failure rates because the synchronization slots are fixed, and when there are straggling workers synchronization cannot proceed.

\section{Implementation and Observations}\label{sec6}

\subsection{Implementation Details}

The time-based redundancy algorithm (shown in Section~\ref{timeAlgo}) is used to implement
{\em multi-point} synchronization in JAMScript. JAMScript is a polyglot language; it brings together C and JavaScript with some additional constructs needed to implement the programming model. The C side implements the {\em worker} side of the programming model and JavaScript side implements the {\em controller} side. This means the C side runs at the lowest level of the system. For instance, embedded devices (e.g., sensors and actuators) will be running the C component. This allows us to target memory constrained devices (planned for the future). The JavaScript running in a NodeJS runtime is responsible for implementing the controllers. The NodeJS based controller can run from low power devices to all the way to cloud servers.

This implementation is achieved in
two stages. In the first stage, a single-point synchronization scheme is realized, where only one controller orchestrates the operation of all workers. At this time, the single-point synchronization is basic and lacks the {\em local scheduler} optimization to minimize the idle time in the workers. Therefore, the local, asynchronous, and synchronous task order is kept the same as that specified in the program. In the second stage, the single-point synchronizer is used to create a multi-point synchronizer by using a hierarchy of controllers. For instance, by placing a controller at \texttt{C1} the operations are synchronized across all the workers in Fig.~\ref{model}. Alternatively, if the controllers are placed at \texttt{F1} and \texttt{F2}, which themselves are coordinated by a super controller at \texttt{C1}, then the workers can be synchronized  in two sub groups: \texttt{G2} and \texttt{G3}. 

\subsection{Example JAMScript Program Snippets}

Listing 1 shows the controller side of the example JAMScript program.
The {\tt trytellid()} function is responsible for calling the synchronous function {\tt tellid()} on the worker nodes. The {\tt trytellid} function is a conditional function that only runs at the fogs.  

That is, when the given JAMScript program is run in a resource configuration shown in Fig.~\ref{model}, the {\tt trytellid} runs only in Fogs {\tt F1} and {\tt F2}. {\tt F1} would synchronize {\tt G2} and {\tt F2} would synchronize {\tt G3}. However, workers in {\tt G2} and {\tt G3} would not be synchronized with each other; thus, presenting multi-point synchronization.

\begin{lstlisting}[language=JavaScript,firstnumber=1, caption=Controller side of the JAMScript example.]
var count = 1;

jcond {
    fogonly: jsys.type == "fog";
}
jsync function getID() {
    return count++;
}
jasync {fogonly} function trytellid() {
    var nodes = tellid();
    console.log("Nodes in the sync group ", nodes);
}
setInterval(function() {
    trytellid();
}, 1000);
\end{lstlisting}\label{listing1}

\begin{lstlisting}[language=C, caption=Worker side of the JAMScript example.]
int getID();
int myid = -1;

jasync trygetid() {
    while(1) {
        jsleep(1000);
        myid = getID();
        printf("MyID %d\n", myid);
    }
}
jsync int tellid() {
    return myid;
}
int main(int argc, char *argv[]) {
    trygetid();
}
\end{lstlisting}\label{listing2}

\subsection{Preliminary Experiments and Discussions}

\begin{figure}[b]
\centering
\includegraphics[height=1.85in]{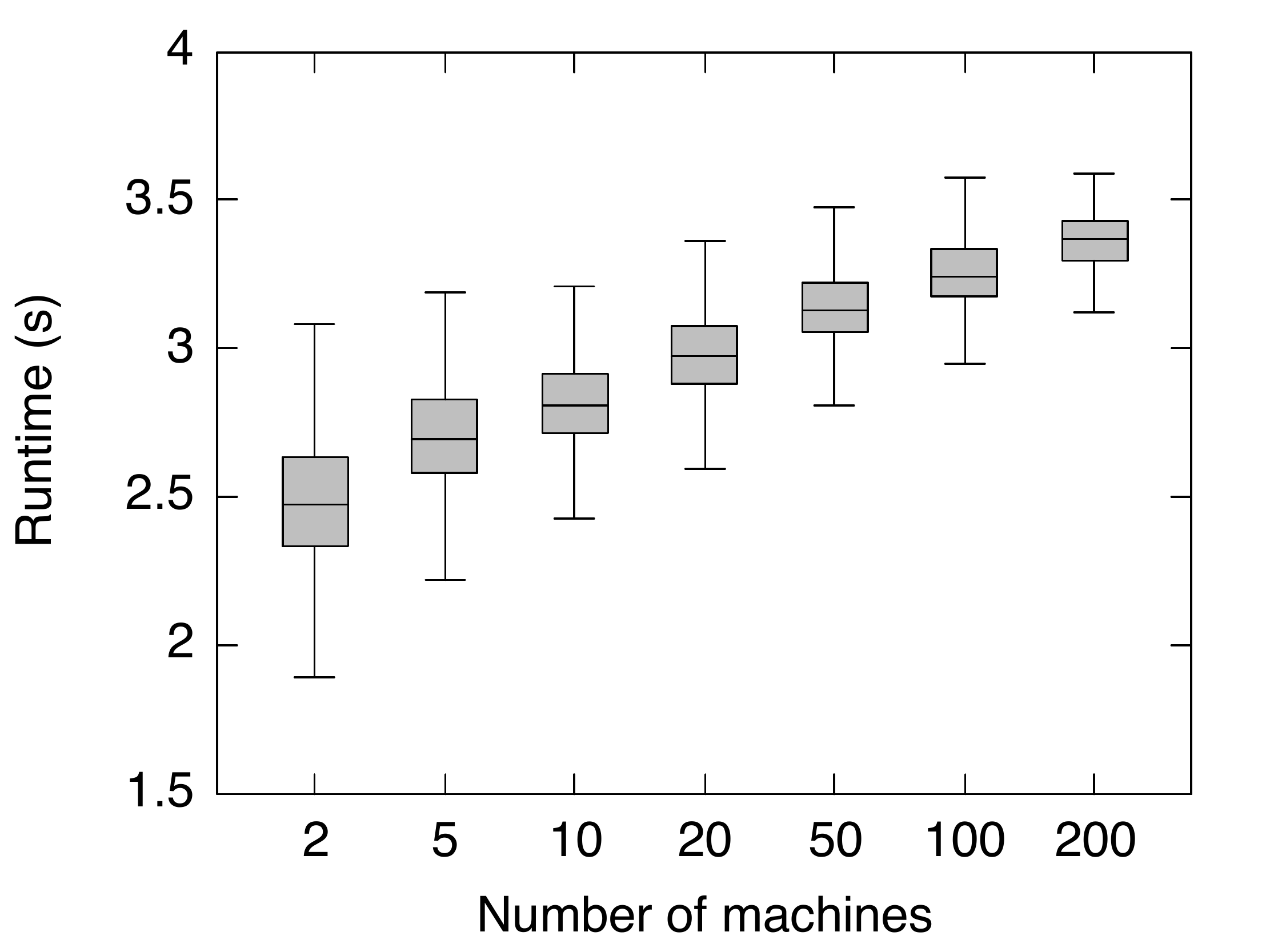}
\caption{Observed execution time(s) for the implementation of the dynamic synchronous scheduling algorithm on JAMScript with the number of workers varying from \textit{2} to \textit{200}.}
\label{implementation}
\end{figure}

Fig.~\ref{implementation} shows the performance of single-point synchronization of a JAMScript program when it is deployed in Docker containers. 

A task graph with synchronous and asynchronous tasks whose execution times on nodes follow a normal distribution with a mean of $1s$ and a variance of $0.25s$ is used. A normal distribution with a mean of $25ms$ and a variance of $10ms$ is used to represent the network delay. The number of nodes is varied from $2$ to $200$.

As part of the future work, an optimized implementation that should reduce the overhead using publish-subscribe and local scheduling will be implemented. It is worthy to note that while JAMScript is used to show the feasibility in implementing the synchronous scheduling algorithms, other controller-centric programming frameworks can be used to implement the algorithms.

\section{Related work}\label{sec7}

Biologically inspired synchronization schemes have been well studied such as pulse coupled oscillator (PCO) synchronization~\cite{hong2005scalable, yadav2017self}. Synchronization in PCO is achieved through a consensus among participating nodes. Assumptions made in PCO are periodic pulse, a mesh connectivity and a zero-communication time. A synchronization scheme combining heartbeat synchronization and a firefly inspired model or overlay networks was proposed in~\cite{babaoglu2007firefly}. The proposed protocol has two main parts, nodes selecting their peer list and processing a flash message received in order to achieve synchronization. 

An emergent broadcast slot synchronization scheme inspired by firefly algorithms was proposed in~\cite{yadav2017self} for Internet of Things (IoT). Each node maintains a time window that it can be awake during a steady state. Nodes in the network go through three states in order to achieve synchronization. The first state is the initialization state where nodes start their random timers and identify their neighbours. The nodes then transition to the  synchronization state, here, the nodes coordinate their  synchronization error tolerance window with their neighbours using a Pulse Coupled Oscillator (PCO) model with a phase advancement function. A node becomes synchronized if its synchronicity is greater than the synchronization threshold. Thereafter, the node moves into the steady duty cycle state. In this phase, nodes only wake up during their  synchronization error tolerance window to exchange messages.

In real time computing systems, synchronization is handled using time-triggered controls where all synchronous activities are executed at some predefined points in time~\cite{vrancic2006synchronization}. Synchronized clocks are used to achieve synchronization in the systems by making each node have a common notion of time. Time synchronization schemes~\cite{guo2016psync, elsts2017temperature} have been developed for IoT to allow devices have a common notion of time. In~\cite{guo2016psync}, a visible light produced by light emitting diodes (LEDs) are used by devices within that vicinity to synchronize. Synchronization is achieved by allowing a number of LEDs to send out binary signals at the same time and at predefined intervals, and devices synchronize when there is a phase transition. In~\cite{elsts2017temperature}, a time synchronization protocol was proposed to mitigate the effect of temperature change on hardware clocks in IoT networks using time-slotted channel hopping. 

Preemptive synchronized scheduling policies~\cite{nemati2009multiprocessor, zeng2014scheduling} have been proposed for synchronization tasks that share common attributes. In~\cite{nemati2009multiprocessor}, a preemptive synchronized scheduling policy was proposed for synchronization tasks that share same mutual exclusion resources under homogeneous processors. Tasks are allowed to share resources based on a mutual exclusion policy. Tasks that share mutually exclusive logical resources are grouped into the same component and have a local scheduler. Whenever a global scheduler decides on a component to be executed, the local scheduler decides which task gets the resource access. In~\cite{zeng2014scheduling}, tasks are partitioned based on communication, synchronization and mutex. Synchronization tasks having the same priority are expected to arrive at the synchronization point (SP) at the same time and cannot be preempted. The synchronization task with the maximum execution time before the SP is scheduled to run first, while that with the maximum execution time after the SP is scheduled to run immediately after the SP. 
 
The synchronization-aware algorithm proposed in~\cite{lakshmanan2009coordinated} bundles tasks by size or mutex-sharing and attempts to schedule them onto a processor. Bundles that do not directly fit into a processor are put in a separate queue and sorted based on their cost (the penalty of transforming a local mutex into a global mutex) in increasing order. The bundle with the smallest cost is broken down into pieces with the largest piece determined by the size of the largest possible space in the processors. If the process is not successful, a new processor is added, and the partitioning process is repeated. 

In coscheduling~\cite{sobalvarro1998dynamic} and gang scheduling (a stricter form of coscheduling)~\cite{karatza2006scheduling, batat2000gang}, tasks of a parallel job (or gang) are scheduled and executed at the same pre-defined time slot. Time slotting is a fundamental notion in gang scheduling where all or none of the members of the gang are run at any time slot. The Ousterhout matrix (OM) is a model for dividing the resources in the space and time dimensions and mapping jobs to them~\cite{batat2000gang}. The OM for gang scheduling represents processing nodes as columns and time slot as rows. Synchronization is achieved in coscheduling and gang scheduling using busy waiting~\cite{karatza2006scheduling}. \vspace{7px}

\noindent\textbf{Comparison of this Paper and Related Work} \vspace{5px}

This work focuses on task and controller driven synchronization. There is a combination of local, asynchronous and synchronous tasks that need to be scheduled unlike the firefly-based (PCO) synchronization schemes ~\cite{hong2005scalable, yadav2017self, babaoglu2007firefly} where synchronization is on a single flash message. The proposed algorithms guarantee atomic $QoS_{ync}$ through quorum checking. That is, the execution of a synchronous task cannot proceed unless the synchronicity conditions are met. Whereas in firefly-based (PCO) synchronization, eventual synchronization is sought for. Preemption is not allowed in our scheme unlike in co- and gang scheduling ~\cite{karatza2006scheduling, sobalvarro1998dynamic} where tasks that have already started execution can be called back. The proposed synchronization schemes are different from barrier synchronization in that there is no need to wait indefinitely for stragglers as synchronization can proceed once the conditions for quorum are met or the necessary precautions taken if quorum fails. The proposed schemes have dynamic synchronization points that can be adjusted based on the availabilities of workers unlike time slotted synchronization where the synchronization points are predefined.

This work extends beyond  time synchronization ~\cite{guo2016psync, elsts2017temperature} in that nodes need not only  have the same notion of time, but there is a need for the right availability of nodes before executing a synchronous task. Tasks are not bundled to be scheduled together on the same processor in our scheme as in ~\cite{lakshmanan2009coordinated}, rather, the same set of tasks are run across all participating nodes. Node disconnections, leaving and joining are catered for by introducing quorum checking to verify the availability and participation of nodes before executing a synchronous task. Fault tolerance is achieved by introducing time and component redundancy.

\section{Conclusions and Future Work}\label{sec8}

The need for synchronization in edge-enabled IoT is motivated with application scenarios and use cases. A system model is developed for mapping applications with and without synchronization requirements onto an edge-enabled IoT system. Two dynamic scheduling algorithms 
are proposed for mapping applications with different quorum requirements. The applications can have 
synchronous, asynchronous and local tasks in them. The tasks are mapped onto a group of single-threaded heterogeneous nodes that can possibly disconnect from each other. 
The challenge for the synchronization algorithms is to keep synchrony between the task executions despite disconnections. 

The synchronization scheduling algorithms use two ideas: time-based and component-based redundancies. It was observed that time-based redundancy is suitable for applications where repeating the task
executions is acceptable. Whereas the component-based redundancy is needed for applications that cannot wait for task re-executions. Experiments are conducted to evaluate the performance of the proposed algorithms and to explore the different trade-offs between the two approaches. 

Updating the controller with the progress in task executions is important for synchronization scheduling. 
It was found that using a publish-subscribe update scheme reduces the communication load on controllers; thus, effectively reducing the overall execution times of the synchronous tasks. It was observed that increasing the level of redundancy for component-based redundancy decreases the runtime and reduces the percentage of sync task failures. Likewise, the prediction accuracy of the finish time of tasks on the workers has a significant impact on the runtime and synchronization task failure. The proposed algorithms have shorter runtimes than barrier synchronization and have less synchronization task failures when compared to time slotted synchronization.

One area of future work is applying machine learning for completion time prediction and incorporating that into synchronization scheduling. Another is to handle device mobility across the fogs. In particular, with vehicular clouds, it is possible to have vehicles joining and leaving different fog zones as they travel. The synchronization scheduler needs to control the vehicle to fog associations to minimize synchronization task failures due to mobility.

\bibliographystyle{IEEEtran}
\bibliography{paper}

\begin{thebibliography}{10}
\providecommand{\url}[1]{#1}
\csname url@samestyle\endcsname
\providecommand{\newblock}{\relax}
\providecommand{\bibinfo}[2]{#2}
\providecommand{\BIBentrySTDinterwordspacing}{\spaceskip=0pt\relax}
\providecommand{\BIBentryALTinterwordstretchfactor}{4}
\providecommand{\BIBentryALTinterwordspacing}{\spaceskip=\fontdimen2\font plus
\BIBentryALTinterwordstretchfactor\fontdimen3\font minus
  \fontdimen4\font\relax}
\providecommand{\BIBforeignlanguage}[2]{{%
\expandafter\ifx\csname l@#1\endcsname\relax
\typeout{** WARNING: IEEEtran.bst: No hyphenation pattern has been}%
\typeout{** loaded for the language `#1'. Using the pattern for}%
\typeout{** the default language instead.}%
\else
\language=\csname l@#1\endcsname
\fi
#2}}
\providecommand{\BIBdecl}{\relax}
\BIBdecl

\bibitem{abedin2015fog}
S.~F. Abedin, M.~G.~R. Alam, N.~H. Tran, and C.~S. Hong, ``A fog based system
  model for cooperative iot node pairing using matching theory,'' in
  \emph{Network Operations and Management Symposium (APNOMS), 2015 17th
  Asia-Pacific}.\hskip 1em plus 0.5em minus 0.4em\relax IEEE, 2015, pp.
  309--314.

\bibitem{wu2011clock}
Y.~C. Wu, Q.~Chaudhari, and E.~Serpedin, ``Clock synchronization of wireless
  sensor networks,'' \emph{IEEE Signal Processing Magazine}, vol.~28, no.~1,
  pp. 124--138, 2011.

\bibitem{guo2016psync}
X.~Guo, M.~Mohammad, S.~Saha, M.~C. Chan, S.~Gilbert, and D.~Leong, ``Psync:
  Visible light-based time synchronization for internet of things (iot),''
  2016.

\bibitem{lee2014swarm}
E.~A. Lee, B.~Hartmann, J.~Kubiatowicz, T.~S. Rosing, J.~Wawrzynek, D.~Wessel,
  J.~Rabaey, K.~Pister, A.~Sangiovanni-Vincentelli, S.~A. Seshia \emph{et~al.},
  ``The swarm at the edge of the cloud,'' \emph{IEEE Design \& Test}, vol.~31,
  no.~3, pp. 8--20, 2014.

\bibitem{bonomi2012fog}
F.~Bonomi, R.~Milito, J.~Zhu, and S.~Addepalli, ``Fog computing and its role in
  the internet of things,'' in \emph{Proceedings of the first edition of the
  MCC workshop on Mobile cloud computing}.\hskip 1em plus 0.5em minus
  0.4em\relax ACM, 2012, pp. 13--16.

\bibitem{wenger2016programming}
R.~Wenger, X.~Zhu, J.~Krishnamurthy, and M.~Maheswaran, ``A programming
  language and system for heterogeneous cloud of things,'' in
  \emph{Collaboration and Internet Computing (CIC), 2016 IEEE 2nd International
  Conference on}.\hskip 1em plus 0.5em minus 0.4em\relax IEEE, 2016, pp.
  169--177.

\bibitem{karatza2006scheduling}
H.~D. Karatza, ``Scheduling gangs in a distributed system,''
  \emph{International Journal of Simulation: Systems, Science Technology, UK
  Simulation Society}, vol.~7, no.~1, pp. 15--22, 2006.

\bibitem{sobalvarro1998dynamic}
P.~G. Sobalvarro, S.~Pakin, W.~E. Weihl, and A.~A. Chien, ``Dynamic
  coscheduling on workstation clusters,'' in \emph{Workshop on Job Scheduling
  Strategies for Parallel Processing}.\hskip 1em plus 0.5em minus 0.4em\relax
  Springer, 1998, pp. 231--256.

\bibitem{valiant1990bridging}
L.~G. Valiant, ``A bridging model for parallel computation,''
  \emph{Communications of the ACM}, vol.~33, no.~8, pp. 103--111, 1990.

\bibitem{khayyat2013mizan}
Z.~Khayyat, K.~Awara, A.~Alonazi, H.~Jamjoom, D.~Williams, and P.~Kalnis,
  ``Mizan: a system for dynamic load balancing in large-scale graph
  processing,'' in \emph{Proceedings of the 8th ACM European Conference on
  Computer Systems}.\hskip 1em plus 0.5em minus 0.4em\relax ACM, 2013, pp.
  169--182.

\bibitem{jakovits2012stratus}
P.~Jakovits, S.~N. Srirama, and I.~Kromonov, ``Stratus: A distributed computing
  framework for scientific simulations on the cloud,'' in \emph{2012 IEEE 14th
  International Conference on High Performance Computing and Communication \&
  2012 IEEE 9th International Conference on Embedded Software and
  Systems}.\hskip 1em plus 0.5em minus 0.4em\relax IEEE, 2012, pp. 1053--1059.

\bibitem{vogli2015fast}
E.~Vogli, G.~Ribezzo, L.~A. Grieco, and G.~Boggia, ``Fast join and
  synchronization schema in the ieee 802.15. 4e mac,'' in \emph{2015 IEEE
  Wireless Communications and Networking Conference Workshops (WCNCW)}.\hskip
  1em plus 0.5em minus 0.4em\relax IEEE, 2015, pp. 85--90.

\bibitem{ozil2007time}
I.~Ozil and D.~R. Brown, ``Time-slotted round-trip carrier synchronization,''
  in \emph{2007 Conference Record of the Forty-First Asilomar Conference on
  Signals, Systems and Computers}.\hskip 1em plus 0.5em minus 0.4em\relax IEEE,
  2007, pp. 1781--1785.

\bibitem{liu2015}
F.~Liu and W.~Guo, ``The design and implementation of mina-based smart home
  data synchronization system,'' in \emph{2015 Fifth International Conference
  on Instrumentation and Measurement, Computer, Communication and Control
  (IMCCC)}, Sept 2015, pp. 1612--1616.

\bibitem{chalupnivckova2014use}
H.~Chalupn{\'\i}{\v{c}}kov{\'a}, P.~Bahensk{\`y}, V.~S{\`y}kora, and
  D.~Heralov{\'a}, ``The use of drones in air cargo transportation,''
  \emph{Economy \& Society \& Environment}, vol. 4509, pp. 1--6, 2014.

\bibitem{glancy2015autonomous}
D.~J. Glancy, ``Autonomous and automated and connected cars-oh my: first
  generation autonomous cars in the legal ecosystem,'' \emph{Minn. JL Sci. \&
  Tech.}, vol.~16, p. 619, 2015.

\bibitem{hasan2011intelligent}
M.~N. Hasan, S.~Didar-Al-Alam, and S.~R. Huq, ``Intelligent car control for a
  smart car,'' \emph{International Journal of Computer Applications}, vol.~14,
  no.~3, pp. 15--19, 2011.

\bibitem{ademaj2007time}
A.~Ademaj and H.~Kopetz, ``Time-triggered ethernet and ieee 1588 clock
  synchronization,'' in \emph{Precision Clock Synchronization for Measurement,
  Control and Communication, 2007. ISPCS 2007. IEEE International Symposium
  on}.\hskip 1em plus 0.5em minus 0.4em\relax IEEE, 2007, pp. 41--43.

\bibitem{gautam2012review}
A.~Gautam and S.~Mohan, ``A review of research in multi-robot systems,'' in
  \emph{Industrial and Information Systems (ICIIS), 2012 7th IEEE International
  Conference on}.\hskip 1em plus 0.5em minus 0.4em\relax IEEE, 2012, pp. 1--5.

\bibitem{kreutz2015software}
D.~Kreutz, F.~M. Ramos, P.~E. Verissimo, C.~E. Rothenberg, S.~Azodolmolky, and
  S.~Uhlig, ``Software-defined networking: A comprehensive survey,''
  \emph{Proceedings of the IEEE}, vol. 103, no.~1, pp. 14--76, 2015.

\bibitem{gracia2013actively}
R.~Gracia-Tinedo, M.~S. Artigas, A.~Moreno-Martinez, C.~Cotes, and P.~G. Lopez,
  ``Actively measuring personal cloud storage,'' in \emph{Cloud Computing
  (CLOUD), 2013 IEEE Sixth International Conference on}.\hskip 1em plus 0.5em
  minus 0.4em\relax IEEE, 2013, pp. 301--308.

\bibitem{liu2010towards}
S.~Liu, Y.~Liu, L.~M. Ni, J.~Fan, and M.~Li, ``Towards mobility-based
  clustering,'' in \emph{Proceedings of the 16th ACM SIGKDD international
  conference on Knowledge discovery and data mining}.\hskip 1em plus 0.5em
  minus 0.4em\relax ACM, 2010, pp. 919--928.

\bibitem{hong2005scalable}
Y.-W. Hong and A.~Scaglione, ``A scalable synchronization protocol for large
  scale sensor networks and its applications,'' \emph{IEEE Journal on Selected
  Areas in Communications}, vol.~23, no.~5, pp. 1085--1099, 2005.

\bibitem{yadav2017self}
P.~Yadav, J.~A. McCann, and T.~Pereira, ``Self-synchronization in duty-cycled
  internet of things (iot) applications,'' \emph{IEEE Internet of Things
  Journal}, vol.~4, no.~6, pp. 2058--2069, 2017.

\bibitem{babaoglu2007firefly}
O.~Babaoglu, T.~Binci, M.~Jelasity, and A.~Montresor, ``Firefly-inspired
  heartbeat synchronization in overlay networks,'' in \emph{Self-Adaptive and
  Self-Organizing Systems, 2007. SASO'07. First International Conference
  on}.\hskip 1em plus 0.5em minus 0.4em\relax IEEE, 2007, pp. 77--86.

\bibitem{vrancic2006synchronization}
A.~Vrancic, ``Synchronization of distributed systems,'' Sep.~26 2006, uS Patent
  7,114,091.

\bibitem{elsts2017temperature}
A.~Elsts, X.~Fafoutis, S.~Duquennoy, G.~Oikonomou, R.~J. Piechocki, and
  I.~Craddock, ``Temperature-resilient time synchronization for the internet of
  things,'' \emph{IEEE Transactions on Industrial Informatics}, 2017.

\bibitem{nemati2009multiprocessor}
F.~Nemati, M.~Behnam, and T.~Nolte, ``Multiprocessor synchronization and
  hierarchical scheduling,'' in \emph{Parallel Processing Workshops, 2009.
  ICPPW'09. International Conference on}.\hskip 1em plus 0.5em minus
  0.4em\relax IEEE, 2009, pp. 58--64.

\bibitem{zeng2014scheduling}
S.~Zeng, B.~He, and J.~Jiang, ``A scheduling algorithm for synchronization task
  in embedded multicore systems⋆,'' \emph{Journal of Computational
  Information Systems}, vol.~10, no.~19, pp. 8531--8541, 2014.

\bibitem{lakshmanan2009coordinated}
K.~Lakshmanan, D.~de~Niz, and R.~Rajkumar, ``Coordinated task scheduling,
  allocation and synchronization on multiprocessors,'' in \emph{Real-Time
  Systems Symposium, 2009, RTSS 2009. 30th IEEE}.\hskip 1em plus 0.5em minus
  0.4em\relax IEEE, 2009, pp. 469--478.

\bibitem{batat2000gang}
A.~Batat and D.~G. Feitelson, ``Gang scheduling with memory considerations,''
  in \emph{Parallel and Distributed Processing Symposium, 2000. IPDPS 2000.
  Proceedings. 14th International}.\hskip 1em plus 0.5em minus 0.4em\relax
  IEEE, 2000, pp. 109--114.

\end{thebibliography}

\begin{IEEEbiography}[{\includegraphics[width=1in,height=1.25in,clip,keepaspectratio]{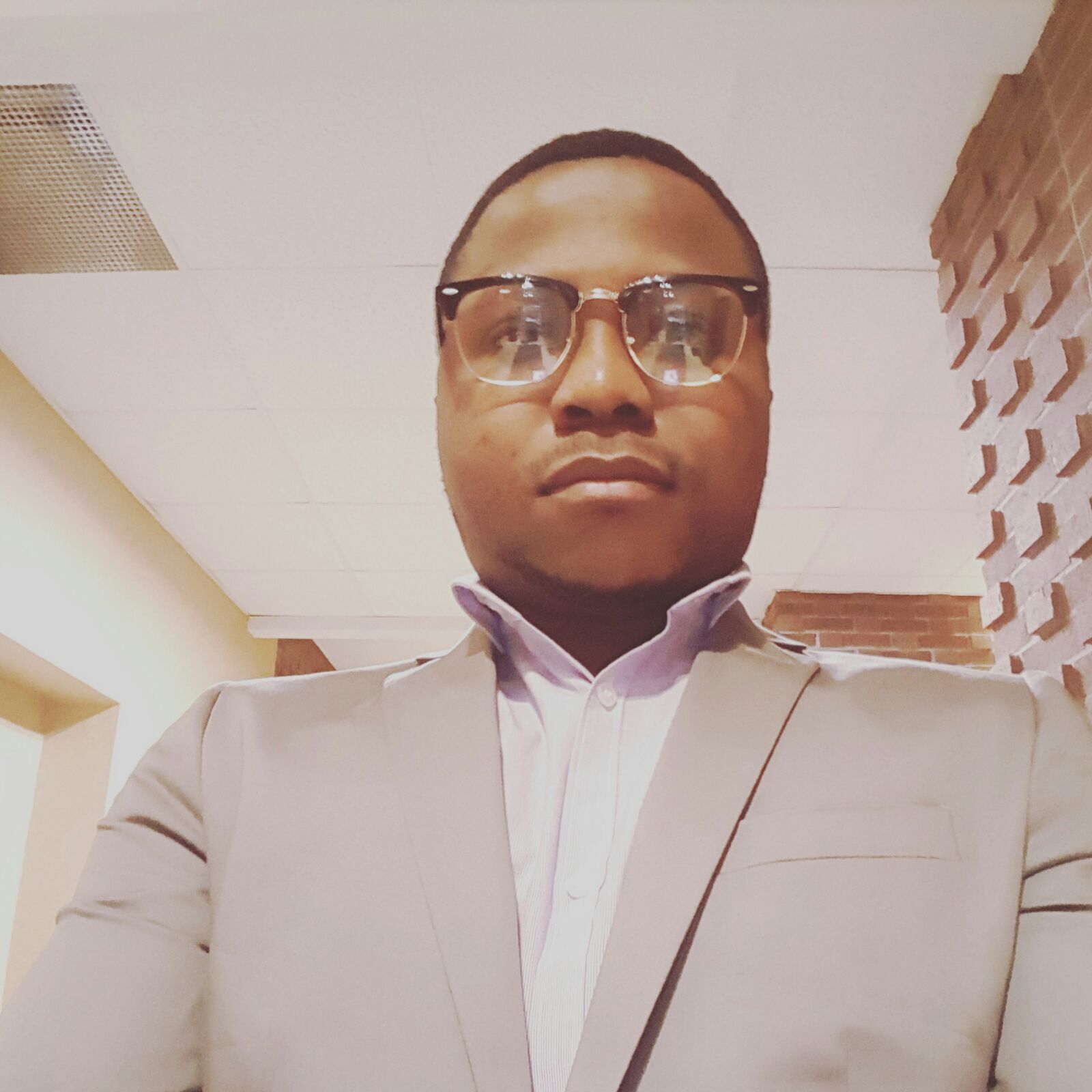}}]{Richard Olaniyan}
 is a PhD student in the School of Computer Science, McGill University, Montreal, Canada being sponsored by the Presidential Scholarship Scheme of the Nigerian Government/Petroleum Technology Development Fund (PTDF) Nigeria. He received his MSc degree in Computer Science at the University of Edinburgh, United Kingdom in 2015. He Received his BSc degree in Computer Engineering from Obafemi Awolowo University, Ile-Ife, Nigeria in 2011, where he graduated as the best student in the department bagging two awards. His research interests include synchronization and scheduling in clouds, clusters, fog computing, edge computing, vehicular clouds and computing models. 
\end{IEEEbiography}

\begin{IEEEbiography}[{\includegraphics[width=1in,height=1.25in,clip,keepaspectratio]{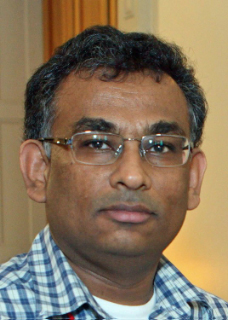}}]{Muthucumaru Maheswaran}
 is an associate professor in the School of Computer Science at McGill University. He got a PhD in Electrical and Computer Engineering from Purdue University, West Lafayette and a BScEng degree in Electrical and Electronic Engineering from the University of Peradeniya, Sri Lanka. He has researched various issues in scheduling, trust management, and scalable resource discovery mechanisms in Clouds and Grids. Many papers he co-authored in resource management systems have been highly cited by other researchers in the area. Recently, his research has focused in security, resource management, and programming frameworks for Cloud of Things. He has supervised the completion of 8 PhD theses in the above areas. He has published more than 120 technical papers in major journal, conferences, and workshops. He holds a US patent in wide-area content routing.
\end{IEEEbiography}
\vfill
\end{document}